\documentclass{article}
\usepackage{geometry}                % See geometry.pdf to learn the layout options. There are lots.
\usepackage{graphicx}
\usepackage{stmaryrd}
\usepackage{amssymb}
\usepackage{amsmath, cancel, centernot }
\usepackage[mathscr]{eucal}
\usepackage{mathtools}
\usepackage{epstopdf}
\title{$p$-adic Distance, Finite Precision and Emergent Superdeterminism: A Number-Theoretic Consistent-Histories Approach to Local Quantum Realism}

\author{T.N.Palmer\\ Department of Physics, University of Oxford, UK\\
tim.palmer@physics.ox.ac.uk}
\date{September 20th, 2016}                                           % Activate to display a given date or no date
\makeatletter
\newcommand\be{\@ifstar{\[}{\begin{equation}}}
\newcommand\ee{\@ifstar{\]}{\end{equation}}}
\newcommand\bp{\begin{pmatrix}}
\newcommand\ep{\end{pmatrix}}

\makeatother
\begin{document}
\bibliographystyle{plain}
\maketitle
\begin{abstract}
Although the notion of superdeterminism can, in principle, account for the violation of the Bell inequalities, this potential explanation has been roundly rejected by the quantum foundations community. The arguments for rejection, one of the most substantive coming from Bell himself, are critically reviewed. In particular, analysis of Bell's argument reveals an implicit unwarranted assumption: that the Euclidean metric is the appropriate yardstick for measuring distances in state space. Bell's argument is largely negated if this yardstick is instead based on the alternative $p$-adic metric. Such a metric, common in number theory, arises naturally when describing chaotic systems which evolve precisely on self-similar invariant sets in their state space. A locally-causal realistic model of quantum entanglement is developed, based on the premise that the laws of physics ultimately derive from an invariant-set geometry in the state space of a deterministic quasi-cyclic mono-universe. This dynamically invariant self-similar subset is locally homeomorphic to $\mathbb Z_2 \times \mathbb R$ where $\mathbb Z_2$ denotes the set of 2-adic integers and $\mathbb R$ denotes a state-space trajectory, or history. Based on this, the notion of a complex Hilbert vector is reinterpreted in terms of an uncertain selection from a finite sample space of states, leading to a novel form of `consistent histories' based on number-theoretic properties of the transcendental cosine function. For example, for a Mach-Zehnder experiment with phase angle $\phi$, histories where $\cos \phi$ and $\phi/\pi$ are describable by a finite number of bits are, by number theory, almost always mutually incompatible, all supplementary variables being equal. This leads to novel realistic interpretations of position/momentum non-commutativity, EPR, the Bell Theorem and the Tsirelson bound. In this inherently holistic theory - neither conspiratorial, retrocausal, fine tuned nor nonlocal - superdeterminism is not invoked by fiat but is emergent from these `consistent histories' number-theoretic constraints.  Because of finite experimental precision,  experimenters have no direct control on whether $\cos \phi$ is finitely describable or not. Hence, Bell inequalities are violated without constraining experimenter free will. Quantum decoherence is described by chaotic riddled-basin dynamics, leading to a natural clustering of trajectories ('measurement eigenstates') on the invariant set. The algebraically closed complex Hilbert Space and associated Schr\"{o}dinger/Dirac equation arise in the singular limit when a fractal parameter $N$ goes to infinity. Invariant set theory provides new perspectives on many of the contemporary problems at the interface of quantum and gravitational physics, and, if correct, may signal the end of particle physics beyond the Standard Model.

\end{abstract}
\newpage
\tableofcontents
\newpage

\section{Introduction}

The recent experiments by Shalm et al \cite{Shalm} on entangled photon pairs appear to have nailed the coffin on local realism, i.e. have comprehensively ruled out putative theories of quantum physics which are both locally causal and deterministic. However, this conclusion is predicated on an assumption which is impossible to test with this type of experiment: that the experimental measurement settings are physically independent of any of the properties of the entangled particles. This is variously referred to as the `Free Choice' or `Free Will' assumption, the `No Conspiracy' assumption, or, more neutrally, the `Measurement Independence' assumption. A theory in which this assumption is violated is referred to as `superdeterministic'. Most physicists who work in the field of either quantum foundations or quantum information theory feel that this assumption is self evident and are consequently almost unanimous in rejecting superdeterministic approaches to fundamental physics. Reasons for rejection include implausible conspiracies, an inability to do science, violation of experimenter free will, or unacceptable fine tuning. In this paper I will explain why I believe the arguments against superdeterminism are unconvincing, and will propose a new theory of quantum physics - invariant set theory - which is not superdeterministic by fiat, but from which plausible superdeterminism is emergent. Although locally causal and realistic, the model violates the Bell inequalities as does quantum theory. It has none of the objectionable properties typically associated with a superdeterministic theory. 

To make such a claim I need some `new meat':
\begin{itemize}

\item Firstly and most importantly, the developments in this paper are based around the use of the $p$-adic metric. Number theory provides us with precisely two norm-based metrics: the Euclidean and the $p$-adic \cite{Katok}. It is proposed that both play central but distinct roles in physics: the former in space-time, the latter in state space. Use of the $p$-adic metric in state space is consistent with a physical theory where states evolve \emph{precisely} on certain self-similar dynamically invariant geometries in state space. This can arise if the corresponding laws of physics are ultimately derived from such state-space geometries. 

\item Secondly, a representation of the multiplicative group of $2^N$th roots of unity is described in terms of cyclic permutations of $2^N$-element bit strings. Based on this, complex Hilbert vectors (and tensor products) are interpreted in terms of uncertain selections of bits from sets of $2^N$-element sample spaces. 

\item Thirdly, the notion of finite experimental precision is exploited. This, together with the items above, ensures that invariant set theory is consistent with experimenter free will. An interplay between the Euclidean metric in space-time and the $p$-adic metric in state space will be used to demonstrate why the violation of the Bell inequalities is robust and requires no special precision on the part of the experimenter, in setting experimental parameters. The notion of finite precision is actually not so new; it has already been shown to be capable of nullifying the Kochen-Specker theorem \cite{Meyer:1999}. 

\end{itemize}

My primary motivation for this work was less about resolving the Bell Theorem, and more about exploring novel possibilities for synthesising quantum and gravitational physics - ones that contrast with conventional approaches based on quantum field theory - which might help resolve contemporary problems at the interface of quantum and gravitational physics: the dark universe, black hole information loss, the 2nd law of thermodynamics, space-time singularities and so on. Some speculative remarks on these issues are addressed towards the end of the paper. 

The structure of this paper is provided in the contents listing above. Most of the technical discussion is relegated to Appendices. No prior knowledge about the $p$-adic metric is assumed. 

\section{Why is Superdeterminism so Disliked?}
\label{Dislike}

In this Section I review the reasons why superdeterminism has been largely rejected by the quantum foundations community and argue why I find these reasons unconvincing. In laying out these arguments, my purpose is not to change sceptical minds, but rather to try to persuade readers that might otherwise be completely dismissive about the concept of superdeterminism, to read on.  

Following the CHSH version of the Bell Theorem, Alice and Bob make measurements on some entangled physical system. The measurement apparatus available to each experimenter has two possible settings, referred to as 0 and 1. For each setting, the measurement has two possible outcomes, also 0 and 1. If $x$ and $y$ refer to the settings, and $a$ and $b$ to the outcomes, then an experiment over multiple measurements determines conditional frequencies of occurrence $p(a,b|x,y)$. A realistic theory attempts to explain these frequencies by assuming that $a$ and $b$ are determined both by $x$ and $y$ and some supplementary information, represented by the generic variable $\lambda$, over which the experimenters do not have control. In this way we can write
\begin{equation}
\label{probs}
p(a,b|x,y) = \sum_{\lambda} p(a,b|x,y, \lambda) p(\lambda|x,y)
\end{equation}
This equation has been written in a general probabilistic form, commonplace in modern accounts of the Bell Theorem (e.g. \cite{Brunner}). This allows for probabilistic as well as deterministic models. In the discussions below, I will be discussing a strictly deterministic model, so that either $p(a,b|x,y, \lambda)=1$ or $p(a,b|x,y, \lambda)=0$. Indeed, as discussed in Section \ref{Dynamics}, determinism is critically important in this model, unlike in more conventional hidden-variable models. By contrast, $p(\lambda| x,y)$ denotes some non-trivial probability distribution, defined by frequentism on finite sample spaces $\Lambda_{xy}$. 

The Measurement Independence assumption states that $\lambda$ is not correlated with $x$ and $y$, so that
\begin{equation}
\label{mi}
p(\lambda|x,y)=p(\lambda)
\end{equation}
or equivalently that $\Lambda_{xy}=\Lambda$, independent of $x$ and $y$. As first recognised by Bell himself, if (\ref{mi}) is violated in some putative hidden-variable theory, then this theory will not necessarily violate the Bell inequality, even if the theory is locally causal and deterministic. A theory in which (\ref{mi}) does not hold is said to be superdeterministic. 

However, with one notable exception \cite{tHooft:2015b, tHooft:2015} discussed in Section \ref{tHooft}, contemporary researchers are largely unequivocal about the correctness of (\ref{mi}). Wiseman and Cavalcanti \cite{WisemanCavalcanti} summarise the prevailing view:
\begin{quote}
This temptation [to reject (\ref{mi})] should vanish if the reader thinks through what it would actually mean to explain away Bell correlations through the real (not just in-principle) failure of free choice. There is no general theory that does this. If such a theory did exist, it would require a grand conspiracy of causal relationships leading to results in precise agreement with quantum mechanics, even though the theory itself would bear no resemblance to quantum mechanics. Moreover, it is hard to imagine why it should only be in Bell experiments that free choices would be significantly influenced by causes relevant also to the observed outcomes. [R]ather, every conclusion based upon observed correlations, scientific or casual, would be meaningless because the observers's method would always be suspect. It seems to us that any such theory would be about as plausible, and appealing, as, belief in ubiquitous alien mind-control.
\end{quote}
This latter point is also emphasised by Ara\'{u}jo \cite{Araujo} who refers to (\ref{mi}) as the `no-conspiracy' condition:
\begin{quote}
I think [(\ref{mi})] is necessary to even do science, because if it were not possible to probe a physical system independently of its state, we couldn't hope to be able to learn what its actual state is. It would be like trying to find a correlation between smoking and cancer when your sample of patients is chosen by a tobacco company. 
\end{quote}
Shalm et al \cite{Shalm} also highlight the supposedly convoluted nature of physical reality if (\ref{mi}) is violated:
\begin{quote}
If [(\ref{mi})] is not true, then a hidden variable could predict the chosen settings in advance and use that information to produce measurement outcomes that violate a Bell inequality
\end{quote}

Bell \cite{Bellb}, in trying to get to the heart of the issue, advanced what I consider to be the most important argument against superdeterminism. As part of this, Bell shows that experimenter free will is not itself the fundamental issue - though I will discuss human free will in Section \ref{free}. In particular, Bell replaces Alice and Bob with mechanical pseudo-random number generators. For these machines, the output is extraordinarily sensitive to minute variations of the initial conditions. In particular, Bell imagines that $x$ and $y$ are set equal to 0 if the millionth bit of some input variable is a 0 and set equal to 1 if the millionth bit of the same input variable is a 1. That is to say, fixing $x$ fixes something about this input variable.  Bell then writes \cite{Bellb}:
\begin{quote}
But this peculiar piece of information is unlikely to be the vital piece for any distinctively different purpose, i.e. it is otherwise rather useless. With a physical shuffling machine, we are unable to perform the analysis to the point of saying just what peculiar feature of the input is remembered by the output. But we can quite reasonably assume that it is not relevant for other purposes. In this sense the output of such a device is indeed a sufficiently free variable for the purpose at hand. For this purpose the assumption [(\ref{mi})] is then true enough, and the [Bell] theorem follows.
\end{quote}

These arguments may seem compelling. However, below I attempt to provide critiques of all of them, leaving Bell's argument for last. 

In my view, it seems illogical to say that if a superdeterministic theory did exist, it would require a grand conspiracy of causal relationships.  The word `conspiracy' implies something secret or covert. If a plausible superdeterministic theory of quantum physics were somehow to be formulated, then by definition the causal structure of the universe would be laid open for all to see and understand. There would be no conspiracies by definition. Referring to (\ref{mi}) as a `no-conspiracy' condition is therefore prejudicial.  

Moreover, if a plausible superdeterministic theory were able to be formulated, then it also seems to me unlikely that it is only in Bell experiments that `free choices would be influenced by causes relevant also to the observed outcomes'. Rather, this situation would become the norm where all non-classical measurements are made, i.e. where the quantum mechanical observables are non-commutative. Far from being special, this situation may be completely ubiquitous in quantum physics. Does that make the problem worse? Far from it, as I attempt to show below. 

The notion that if a superdeterministic theory did exist, its results would have to be `in agreement with quantum mechanics even though the theory would bear no resemblance to quantum mechanics', could be a merit rather than a drawback. Results from General Relativity agree well with Newtonian Gravity in the appropriate limit, even though General Relativity bears no resemblance to Newtonian theory. Given the ongoing problems synthesising quantum and gravitational physics, some might view the formulation of a deterministic locally causal (\emph{a fortiori} geometric) theory, whose results agree with quantum theory but which bears no resemblance to quantum theory, to be something that is sorely needed! This is my view, at least. 

And before worrying about `alien mind control' I argue in Section \ref{Gene} that by rejecting superdeterminism \emph{a priori}, we may, ironically, have been subverted in our thinking about the Bell Theorem by something just as insidious but rather closer to home - gene control!

Shalm et al's notion that a hidden variable could somehow predict the chosen settings in advance and use that information to ensure measurement statistics are quantum theoretic, is rather anthropomorphic. It conjures up a picture of a Laplacian demon, furiously computing the future before it happens, and then rearranging things to ensure the quantum theoretic result always obtains. If this is how the world works, it would indeed be bizarre. However, if the demon could predict the chosen settings in advance, then ordinary determinism (never mind superdeterminism) would be in deep trouble. Suppose $\lambda$ is a set of Cauchy data on a spacelike hypersurface in the past of the event where Alice chooses $x$. If the demon could use $\lambda$ to predict Alice's choice, and somehow chalk it up for her to see, then Alice would merely have to choose differently to the predicted choice to create a logical paradox. There is no paradox because, even if the laws of physics are deterministic, they are not predictably so. From a practical perspective one could put it like this: if the laws of physics are sufficiently chaotic, any reliable prediction would require a computational capability exceeding that of the whole universe. Since our putative demon would necessarily be a subset of the universe, his limited computational capability prevents such reliable prediction. These ideas can be expressed more formally in terms of the G\"{o}del-Turing theorem: there are deterministic systems whose properties are generically not algorithmic at all. Fractals, the limiting states at $t=\infty$ arising from certain chaotic systems and to which we return below, provide a beautiful geometric manifestation of the G\"{o}del-Turing theorem \cite{Blum} \cite{Dube:1993}. (For this reason, there is actually no need to keep the supplementary variables $\lambda$ `hidden' in a model which encodes the unpredictability of the real world.) In summary, Shalm et al's argument against superdeterminism assumes properties which such a theory need not have. 

The argument that we can't do science when the system being measured is not fully independent of the system that performs the measurement is reminiscent of the argument that it is not possible to do objective science on the topic of consciousness, since  we, the investigators, are conscious beings and therefore not fully independent of the topic we investigate. However, given the extraordinary advances in the science of consciousness in the last decade or so (e.g. using Functional MRI), this argument does not appear to be sound. In the case of consciousness, it merely indicates that we have to be careful when designing and analysing experiments which probe our and other animals' consciousness, and I would say the same about analysis of experiments which probe quantum physics. Fortunately, there is a way to analyse data without potential self-referential limitations. That is to say, nature has given us the capability to study causal relationships where the system and the measuring apparatus have become decoupled. It is the classical limit. Hence, if we want to do science where we are sure that classical notions of measurement independence hold, then we must do it in the classical limit (where observables commute). In any case, to say that we can't do science when not in the classical limit is countered explicitly by the theory discussed below, which are being based on (what I think are) sound scientific principles.  

In my view, the most important argument against superdeterminism was provided by Bell and this needs more extensive analysis. In Bell's example, the number `one million' is there to emphasise that the output of the pseudo-random number machine depends on something arbitrarily small (the logic works \emph{a fortiori} if we replace `one million' by `one trillion'). Since the supplementary variables $\lambda$ in principle describe all that is going on in the universe other than these bits - the moons of Jupiter for example\footnote{In \cite{Bellb} Bell notes: `In this matter of causality, it is a great inconvenience that the real world is given to us only once. We cannot know what would have happened had something been different. We cannot repeat an experiment changing just one variable; the hands of the clock will have moved and the moons of Jupiter.' A perturbation which changes just one variable is therefore a perturbation in state space, not in space-time.} - then Bell is drawing attention to our intuition that surely the moons of Jupiter couldn't care less whether that millionth bit here on Earth was a 0 or a 1. Implicit in Bell's argument is that if for some strange reason you think that they do care about the millionth digit, then surely they won't care about the trillionth digit (and son on)! Hence, if we posit some superdeterministic theory of quantum physics, then we need some argument to explain why the moons do `care', no matter how small the effect that sets $x$ and $y$. Maybe we could suppose that, for some theoretical reason, parts of the state space of the universe contains gaps or lacunae: regions where, for some theoretical reason, no space-time trajectories, or `histories', can exist. Maybe by flipping the value of the millionth digit, but keeping fixed the supplementary variables (including the moons of Jupiter), the state of the universe somehow falls into one of these lacunae. However, be this as it may, this doesn't address Bell's concern that in such a theory, the condition of being in a lacuna or otherwise would be sensitive to arbitrarily small perturbations (i.e. to the value of the millionth digit of the random number generator input). Any theory describing this state of affairs would be implausibly fine tuned. We require theories of physics to be structurally stable, with gross properties which are not sensitive to arbitrarily small perturbations. After all, as Bell pointed out, the experimenter's hands tremble when she set the dials on her instruments. A that requires the hands to be completely steady and the values of experimental parameters precise is simply inconsistent with observation. Finely-tuned models, which appear to be implicit in all causal description of the Bell inequalities \cite{WoodSpekkens}, are not acceptable. Game, set and match to Bell? No!

\section{Rethinking the Notion of State-Space Distance}
\label{metric}

There is a subtle but important issue in Bell's argument above: perhaps the single most important single issue in this paper. When we describe some perturbation as `small' what do we mean? Suppose, in a particular run of Bell's pseudo-random number generator, the millionth digit was a 0 (and hence, in the consequent experiment, $x=0$). Label the real world in which these events transpire by $U$, and imagine a counterfactual world $U'$, identical to $U$ in all respects (including the position of the moons of Jupiter) except that the millionth digit was a 1 rather than a 0. Is the distance between $U$ and $U'$ small (say at the time the input variable was being inputted to the pseudo-random number generator)? Certainly in terms of the familiar Euclidean metric, the distance between $U$ and $U'$ can indeed be considered small (and can be made smaller still by making the output of the machine sensitive to the trillionth rather than millionth input bit). 

\begin{figure}
\centering
\includegraphics[scale=0.3]{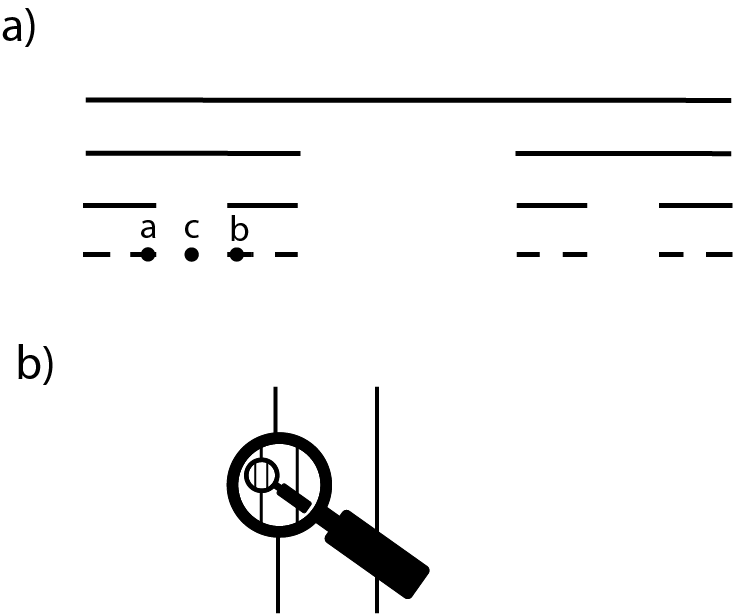}
\caption{a) Some iterates of the Cantor Ternary Set $C(2)$, itself equal to the intersection of all such iterates and homeomorphic to the set of 2-adic integers $\mathbb Z_2 \subset \mathbb Q_2$. The points $a$ and $b$ both belong to $C(2)$ but the point $c$ does not. Both the Euclidean and the 2-adic distance between $a$ and $b$ is small. However, even though the Euclidean distance between $a$ and $c$ is smaller than between $a$ and $b$, the 2-adic distance between $a$ and $c$ is not smaller than between $a$ and $b$. Hence $b$ is 2-adically closer to $a$ than is $c$. b) Two trajectories or histories of the $i$th iterate $C_i(2) \times \mathbb R$ of $C(2) \times \mathbb R$. Zooming into one trajectory reveals two trajectories of $C_{i+1}(2) \times \mathbb R$ and zooming into one of these reveals two trajectories of $C_{i+2}(2) \times \mathbb R$, and so on.}
\label{cantor}
\end{figure}

In assessing the distance between $U$ and $U'$, we are considering distances in state space (recall Footnote 1 above). In physics, we typically use the Euclidean metric to define distances not only in space-time but also in state space. Hence, when the philosopher David Lewis made the following superficially incontrovertible statement in his seminal paper on causation \cite{Lewis}:
\begin{quote}
We may say that one world is closer to actuality than another if the first resembles the actual world more than the second does.
\end{quote}
Lewis is implicitly assuming closeness is synonymous with smallness of the Euclidean metric. Mathematically, the Euclidean metric is ultimately a number-theoretic concept: given numbers $x$ and $y$, the Euclidean distance between them is $|x-y|$ where $|\ldots|$ represents the absolute value (or Euclidean norm). However, in number theory, there does exist another norm-induced metric, called the $p$-adic metric (e.g. \cite{Katok}), where $p$ is typically a prime number. Moreover, by Ostrowski's theorem \cite{Katok}, the $p$-adic metric is the only norm-induced alternative to the Euclidean metric. As will be discussed below (see also Fig \ref{cantor}), Lewis' intuition is false if the notion of closeness is defined $p$-adically. 

Before outlining why, it is worth briefly discussing an analogy which illustrates the importance of using the correct metric. At first sight, Penrose's impossible triangle (Fig 4-7 of \cite{Penrose:2016}) might suggest something incomprehensible about physical space. However, this conclusion is only reached because we visualise the triangle in 2-dimensional Euclidean space and therefore imagine that parts of the triangle that are close in the Euclidean 2D metric, are physically close. However, this is not so. As is well known, the object is constructible in 3-dimensional Euclidean space, whence one pair of sides that appear to meet at a vertex, do not. The ends of these sides are far apart in the more physically relevant Euclidean 3D metric. In the case of the impossible triangle, use of the wrong metric could perhaps lead us into thinking that space is weirder than it really is. 

The $p$-adic metric is very commonly used in number theory:
\begin{quote}

`We [number theorists] tend to work as much $p$-adically as with the reals and complexes nowadays, and in fact it it best to consider all at once.' (Andew Wiles, personal communication 2015.)

\end{quote}
Physicists use real and complex numbers in equal measure, but, so far, not the $p$-adics\footnote{There have been attempts to use $p$-adics in physics, see. e.g. \cite{Volovich, Khrennikov}. However, in these approaches, the $p$-adic metric was used to describe Plankian scales in space-time, and not as a fundamental yardstick in state space. Moreover, so-called $p$-adic integers were not singled out as having special ontological status (see below). The particular use of $p$-adic numbers in this paper is, I believe, new}.  However, if there exist two and only two fundamentally inequivalent metrics in mathematics, why would physics only make use of one of them and completely shun the other? Is there any overriding reason to use the Euclidean metric in state space? I am not aware of one. Are there good physical reasons for using the Euclidean metric in space-time and the $p$-adic metric in state space? I believe there are. 

A brief review of $p$-adic numbers is given in Appendix \ref{padic}. An even briefer introduction is given here. Given a distance function on the rationals $\mathbb Q$, we can complete $\mathbb Q$ based on equivalence classes of Cauchy sequences of rationals. Completion with respect to the Euclidean metric produces the familiar real numbers $\mathbb R$, whilst completion with respect to the $p$-adic metric produces a number system, the field of $p$-adic numbers $\mathbb Q_p$, which is quite different to $\mathbb R$. In my view, the most physical way to understand the differences between these number systems is through their links to geometry. $\mathbb R$ provides the basic building block to analyse Euclidean geometries. By contrast, a subset $\mathbb Z_p$ of $\mathbb Q_p$, known as the set of $p$-adic integers, is homeomorphic to self-similar Cantor sets $C(p)$ with $p$ iterated pieces, i.e. to fractal geometries (see Fig \ref{cantor}a with $p=2$, the value for $p$ used in this paper). We close the circle with physics by noting that fractals play a central role in describing the asymptotic structure of certain deterministic chaotic dynamical systems \cite{Strogatz}, one of the most famous examples being the Lorenz attractor \cite{Lorenz:1963}. Locally, these fractals describe a Cantor set's worth of trajectories, or histories as they are commonly described in quantum theory (I use both words below). A schematic illustration of part of $C(2) \times \mathbb R$ is given in Fig \ref{cantor}b.

To understand the key property of the $p$-adic metric needed to counter Bell's argument against superdeterminism, consider two points $a$ and $b$ which both lie on $C(p)$. Then the $p$-adic distance between $a$ and $b$ can be made as small as one likes by bringing $a$ and $b$ sufficiently close together. (This is related to the fact that $a$ and $b$ can both be represented by elements of $\mathbb Z_p$.) However, consider a third point $c$ which does not lie on $C(p)$, but, let's say, in one of the lacunae between pieces of $C(p)$. Then the $p$-adic distance between $a$ and $c$, or between $b$ and $c$, cannot be made smaller than $p$, no matter how small the Euclidean distance is between $a$ and $c$ or between $b$ and $c$ \footnote{There is a subtlety here. There are many real numbers which have no correspondence in $\mathbb Q_p$ and hence whose $p$-adic distance to the invariant set is undefined (but hence not small). However, $\mathbb Q_p$ can be extended to an algebraically complete field $\mathbb C_p$, isomorphic to the field $\mathbb C$ of complex numbers \cite{Robert}. This suggests that one should start with a state space which is inherently complex, such as that in twistor theory \cite{Penrose:1967}, and then consider an embedded dynamically invariant subset with a topological structure associated with the corresponding ring of integers of $\mathbb C_2$ which is $2$-adically distant from the rest of state space.}  Hence, distance measured using the $p$-adic metric provides a direct and natural way to determine the ontology of putative states of a deterministic system which makes an ontological distinction between those states lying on some fractal invariant subset of state space, and those not lying on this invariant set. The theory that is developed below is predicated on the notion that whilst the Euclidean metric is the metric of choice in space time, the 2-adic metric is the metric of choice in state space. 

What has this got to do with the Bell argument against superdeterminism? Let us suppose that the universe $U$ at any particular time can be considered the state of  a deterministic dynamical system evolving \emph{precisely} on a fractal invariant set $I_U$ \cite{Palmer:2009a}, locally of the form $C(2) \times \mathbb R $ in cosmological state space\footnote{In Appendix \ref{Dirac} we show that the trajectories, or histories, are in fact discretised in the time direction. Hence, strictly, we should write $C(p) \times \mathbb Q$ rather than $C(2) \times \mathbb R$.}   \cite{Palmer:2014}. There are a number of cosmological implications for such a supposition, discussed in Section \ref{Gravity}. Such a model of the universe, although deterministic, is not classical. The states of classical dynamical systems, being based on differential or difference equations of motion, evolve from arbitrary states in state space, and therefore do not generically lie \emph{precisely} on their asymptotic invariant-set attractors except in the (classically unattainable) limit $t=\infty$. As mentioned above, from a $p$-adic perspective, the difference between not lying on $I_U$ and lying on $I_U$ is never small, even for arbitrarily large $t$ (and even if the difference appears very small from a Euclidean perspective). That is to say, from a $p$-adic perspective, lying on $I_U$ precisely at $t=\infty$ is a singular limit \cite{Berry}. The question posed at the beginning of this section can be posed as follows: does the counterfactual world $U'$ where the millionth input bit has been flipped, lie on $I_U$ or not? If it does not, it cannot be considered close to $U$, whether it is the millionth or the trillionth digit that has been flipped. At this stage in the paper we do not have enough information to say whether $U'$ lies on $I_U$ or not. We need to make contact with quantum physics, and complex Hilbert vectors in particular. 

Before concluding this Section, it can be noted that many of the methods of analysis (algebra, calculus, Fourier transforms, and Lie group theory - the bread and butter of theoretical physics) can be applied to the set of $p$-adic numbers \cite{Robert}. In essence, $p$-adic numbers are to fractal geometry as real numbers are to Euclidean geometry. 

\section{Number-Theoretic Consistent Histories and EPR}
\label{interfere}

\begin{figure}
\centering
\includegraphics[scale=0.3]{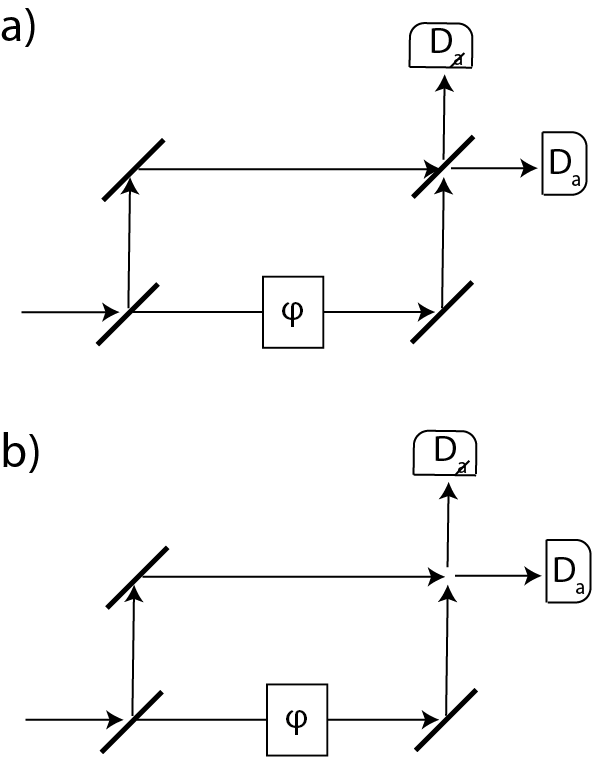}
\caption{a) A momentum experiment. b) A position experiment (obtained by removing the second half-silvered mirror) in the single photon Mach-Zehnder apparatus.}
\label{MachZehnder}
\end{figure}

In this Section, a novel realistic `consistent-histories' interpretation of complex Hilbert vectors is developed. Consider a standard Mach-Zehnder interferometric experiment. Let $x=0$ if the experimenter chooses to perform a momentum measurement (see Fig \ref{MachZehnder}a) and $x=1$ if she chooses to perform a position measurement (see Fig \ref{MachZehnder}b). 
According to quantum theory, the input state vector $|a\rangle$ is transformed by the non-commuting unitary operators
\be
U=\bp \; 1 &1 \\ 1& -1 \; \ep; \ \ \  V=\bp \; 1 &0 \\ 0& e^{i\phi} \; \ep \nonumber
 \ee
When $x=0$, the transformation is given by 
\be
\label{momentum}
|a\rangle \stackrel{UVU}{\mapsto} \cos\frac{\phi}{2}|a\rangle + \sin\frac{\phi}{2} |\cancel a\rangle
\ee
and the probability of detection by the two detectors, is equal to $\cos^2 \phi/2$ and $\sin^2\phi/2$ respectively. When $x=1$, the transformation is given by
\be
\label{position}
|a\rangle \stackrel{VU}{\mapsto} \frac{1}{\sqrt 2}(|a\rangle + e^{i \phi} |\cancel a\rangle)
\ee
and the probability of detection by either of the detectors, and hence the frequency of detection by either detector, is equal to $1/2$. The experimenter is of course free to set $\phi$ and choose between $x=0$ and $x=1$ as she likes. However, necessarily $\phi$ can only be set to finite precision - the experimenter's hands tremble \cite{Bellb} and so cannot have complete control on all the bits that describe $\phi$. This will be important in what is described below. 

Recall that in classical physics, unit vectors in a real Hilbert space provide a natural quantity to represent the uncertain state of a system when - e.g. when describing some future state. For example, let $\mathbf i$ denote a unit vector representing the event $a$: it rains somewhere in London sometime this coming Saturday. The orthogonal unit vector $\mathbf j$ can therefore represent the event $\cancel a$: it rains nowhere in London at any time this coming Saturday. By Pythagoras's theorem, 
\be
\label{realhilbert}
\mathbf{v}(P_a)=\sqrt{P_a}\; \mathbf i+\sqrt{P_{\cancel a}} \;\mathbf j \nonumber
\ee
also has unit norm, for any probability assignment $P_a$ that it will rain somewhere in London this coming Saturday, and $P_{\cancel a}$ that it won't. Hence, in this case, $\mathbf{v}(P_a)$ represents the uncertain state of London's weather this coming Saturday, where the probability that it will rain equals $P_a$. A best estimate of $P_a$ is given by an ensemble weather forecast for Saturday \cite{Palmer:2006}. Such a forecast may typically comprise 50 integrations of a numerical weather forecast model - itself encoding the Navier-Stokes equations -  each individual forecast being made from slightly different starting conditions (consistent with the fact that available weather observations only define the starting conditions imperfectly). Hence $P_a=q/50$ and $P_{\cancel a}=1-q/50$ where $q \le 50$ denotes the number of individual forecasts which predict rain over London on Saturday. Since each weather forecast defines a trajectory of the classical equations of motion, $P_a$ is defined by frequentism over the space of 50 weather trajectories which lie, approximately, on Earth's weather attractor in meteorological state space. We can assume the real weather follows one of these trajectories - though which one is uncertain until we have observed Saturday's actual weather. 

By analogy, we can interpret the real Hilbert vector (\ref{momentum}) in terms of an uncertain selection from some finite string 
\be
\label{string}
\{a_1 \; a_2 \; a_3 \ldots a_{2^N}\}
\ee
where $a_i \in \{a, \cancel a\}$, $N$ is a parameter to be described below, and a fraction $\cos^2 \theta/2$ of the $2^N$ elements are $a\;$s, the remaining $\sin^2 \theta/2\; 2^N$ elements being $\cancel a\;$s. Each element of the string labels a state-space trajectory; $a_i=a$ if a photon is detected by $D_a$, $a_i=\cancel a$ if the photon is detected by $D_{\cancel a}$. By definition, $\cos^2 \theta/2$ and hence $\cos \theta$ is describable by a finite number of bits. Hereafter, I will use the phrase `finitely describable' to mean `describable by a finite number of bits'. It will be convenient to order the bits of (\ref{string}) so that the first $\cos^2 \theta/2$ of the $2^N$ elements are $a\;$s. 

The complex Hilbert vector (\ref{position}) can be interpreted in a similarly realistic way. First let $\theta=\pi/2$ so that the first half of the elements of (\ref{string}) are $a\;$s. Then define a cyclical permutation operator $\zeta$
\be
\label{zeta}
\zeta \{a_1 \; a_2 \; a_3 \ldots a_{2^N}\}= \{a_{2^N} \; a_1 \; a_2 \ldots a_{2^{N-1}}\}
\ee
so that $\zeta^{2^N}$ is equal to the identity operator, i.e. $\zeta$ is a representation of a $2^N$th root of unity. We can therefore write
\be
\label{exp}
\zeta= e^ {2\pi i /2^N}
\ee
as an equivalent expression for the operator $\zeta$. In this way, we will define (\ref{position}) as an uncertain selection of element from the bit string
\be
\zeta^n \{a_1 \; a_2 \; a_3 \ldots a_{2^N}\}
\ee
where $n=2^N \phi/2\pi$. 

In order to avoid disrupting the flow of this Section too much, I summarise below a number of important points to be made about these representations of (\ref{momentum}) and (\ref{position}), leaving more detailed discussion to Sections \ref{measurement}, \ref{Gravity} and Appendix \ref{Hilbert}. 

\begin{itemize}

\item At a more fundamental level, this symbolic labelling of trajectories reflects the fact that each trajectory is attracted to one of two distinct clusters on $I_U$ - labelled $a$ and $\cancel a$ - through a deterministic chaotic (and hence nonlinear) process. These clusters therefore correspond to measurement eigenstates. The parameter $N$ is linked to this chaotic process, reflecting the number of fractal iterates needed to evolve to one cluster or the other. In Section \ref{Gravity} we speculate that this clustering of state-space trajectories is a manifestation of the phenomenon of gravity. 

\item By the discussion above, a momentum measurement on $I_U$ is therefore associated with a finitely describable  $\cos \phi$.  A position measurement on $I_U$ is associated with a finitely describable $\phi/\pi$. 

\item The chaotic procedure which determines which particular element $a_i$ of (\ref{string}) is selected, will be sensitively dependent on the phase angle $\phi$, even though the probability of selecting $a_i$ is independent of $\phi$. The selection procedure is periodic in $\phi$.

\item The bit-string representations of (\ref{momentum}) and (\ref{position}) are examples of bit-string representations of the general single-qubit Hilbert vector $
\cos \frac{\theta}{2}|a\rangle
+e^{i\phi} \sin \frac{\theta}{2} |\cancel{a}\rangle
$ which is invariant under an $SU(2)$ transformation and hence reflects the underlying isotropy of space. As such, the underlying symbolic labelling also reflects this isotropy. and is  spontaneously broken by the clustering procedure mentioned in the bullet above. A general $m$-qubit state is represented by $m$ partially correlated bit strings. 

\end{itemize}

We now come to a number-theoretic result which is central to this paper: 
\\
\\
\textbf{Theorem} \cite{Niven, Jahnel:2005}. It is impossible for both $\phi/\pi$ and $\cos \phi$ to be simultaneously finitely describable unless $\phi=0$, $\pi/2$, $\pi$, $3\pi/2 \ldots$. \\
\\
An elementary proof is given in Appendix \ref{number}. 
\\
\\
A direct consequence of this theorem is the generic non-commutativity of position and momentum measurements - the essence of quantum theory \cite{Griffiths}. Suppose we perform a momentum measurement with the Mach-Zehnder interferometer. Consider the following counterfactual question:  Even though in reality the experiment was a momentum measurement, could it have been a position measurement? In asking this counterfactual question we will imagine that the supplementary variables $\lambda$ - including the moons of Jupiter - are held fixed. Now because in reality the first measurement was a momentum measurement, the set of allowed orientations $\phi$ - corresponding to the set of histories on $I_U$ - are those where $\cos \phi$ is finitely describable. Assuming $\phi$ isn't precisely one of the four exceptions (very unlikely if $2^N \gg 4$), we can scan through all the values of $\phi$ where $\cos \phi$ is finitely describable, and we will \emph{never} find one where $\phi/\pi$ is finitely describable. That is to say, a counterfactual state of the universe corresponding to a position measurement does not and cannot lie on the invariant set $I_U$. Now suppose a second actual measurement with the interferometer was a position measurement, i.e. where  $\phi/\pi$ is finitely describable. Then a second counterfactual momentum measurement, i.e. where $\cos \phi$ is finitely describable, necessarily lies off $I_U$. This is turn means that the order of the two measurements (the first momentum, the second position) could not be reversed - the measurements are non-commutative, consistent with the Uncertainty Principle. 

The finite describability of $\cos \phi$ or $\phi/\pi$ delineates two distinct types of invariant set structure (`momentum structure' and `position structure'). Now as Bell noted, experimenters hands tremble - they only have direct control on the leading bits of $\phi$ and therefore have no direct control, when they set the dials on their instruments, on whether $\cos \phi$ or $\phi/\pi$ is finitely describable. That is to say, whilst the experimenter is completely free to set as many or as few of the bits she can control as she likes, she cannot directly control whether $\cos \phi$ or $\phi/\pi$ is finitely describable. Finite precision is an important consistency condition for this theory: if experimenters could control all the bits of $\phi$, then they would be able to set $\phi/\pi$ to be finitely describable in a situation where a momentum measurement was being made (or $\cos \phi$ to be finitely describable when a position measurement was being made). This inconsistency is reminiscent of the inconsistency described in Section \ref{Dislike} which would arise if the Laplacian demon could predict the future. In both cases the inconsistency is avoided by the consequences of finite precision (in not being able to control whether $\cos \phi$ or $\phi/\pi$ is finitely describable in the first instance, or not being able to predict the future in the second case). 

Also crucial to the consistency of this picture is the use of the $p$-adic metric. In conventional physical theory, it is an irrelevance whether or not $\cos \phi$ is finitely describable: the standard Euclidean distance between a universe where $\cos \phi$ is finitely describable and one where it is not, can be made as small as one likes. Since we require our theories to be structurally stable, a conventional theory where these differences mattered would be considered unacceptably fine tuned because the differences would be destroyed by arbitrarily Euclidean-small noise. However, as discussed, if the counterfactual world $U'$ where $\cos \phi$ is not finitely describable lies off the invariant set, it cannot be $p$-adically close to the real world $U$ where $\cos \phi$ is finitely describable (no matter how close these worlds appear from a Euclidean perspective). As such, one cannot perturb the system off the invariant set with $p$-adic-small amplitude noise. That is to say, the finitely describability of  $\cos \phi$ is structurally stable to $p$-adic noise. 

As Bell has pointed out, the no-go theorems of quantum physics can tolerate a considerable amount of hand trembling. Now hand trembling occurs in space-time, not in state space. We can imagine that as the hand is on the dial, the actual $\phi\;$ realised at any time varies  slightly from one time to the next, where `slightly' is meant in the sense that the Euclidean distance between any two $\phi\;$s is small. However, no amount of such trembling need violate the state-space constraint that all elements of the time series of $\cos \phi$s realised are finitely describable when the hand trembles on the dial. As such, the counterfactual universe $U'$ where $\cos \phi$ is not finitely describable will never be encountered, even momentarily, by any such hand trembling. Bell noted that quantification of hand trembling `would require careful epsilonics'. As this discussion shows, such epsilonics must carefully distinguish space-time epsilonics, where the measure of distance is Euclidean, from state-space epsilons where the measure of distance is $p$-adic. There is a subtle interplay between these different metrics when unravelling these matters. 

Referring to a state-space trajectory as a history, then invariant set theory can be viewed as a realistic 'consistent histories' theory, although at a technical level this is quite different to the conventional notion of Consistent Histories \cite{Griffiths}. For example, here the word 'consistent' has the precise number-theoretic meaning discussed above: if a trajectory with finitely describable $\cos \phi$ lies on $I_U$, then a trajectory with the same supplementary variables $\lambda$ but with finitely describable $\phi/\pi$ is not a consistent history and does not lie on $I_U$. 

This number-theoretic notion of consistency in turn implies that the sample space $\Lambda_{x=0}$ of supplementary variables $\lambda$ associated with finitely describable $\cos \phi$ must be disjoint from the sample space $\Lambda_{x=1}$ of supplementary variables associated with finitely describable $\phi/\pi$. Hence, $\rho(\lambda | x) \ne \rho(\lambda)$, violating (\ref{mi}). A key point is that this is not superdeterminism by fiat (a criticism levelled at other superdeterministic approaches), rather, it is emergent superdeterminism; emergent from this number-theoretic consistent histories approach to complex Hilbert vectors. 

Now if $\phi_1/\pi$ and $\phi_2/\pi$ are both finitely describable, then so is $(\phi_1 - \phi_2)/\pi$. By the theorem of Appendix \ref{number}, ignoring the four exceptions, $\cos (\phi_1-\phi_2)/2$ is not finitely describable. Hence, from the identity
\be
\label{noadd}
\frac{1}{2}(e^{i \phi_1}+e^{i \phi_2})=e^{i \frac{\phi_1+\phi_2}{2}} \cos \frac{\phi_1-\phi_2}{2}
\ee
on the field of complex numbers $\mathbb C$, the multiplicative operators $e^{i \phi}$ in (\ref{exp}) cannot be additive. Fundamentally, it is this property that makes invariant set theory holistic. Without additivity, it is not permissible to break up a system into sub-units. Despite this, it may be computationally convenient to extend the set of angles $\phi/\pi$ from those that are finitely describable to all angles on the whole circle, allowing multiplication and addition of $e^{i\phi}$. Similarly, from a computational point of view we may wish to extend the finitely allowable $\cos \phi$ to the full set of reals on $-1 \le \cos \phi \le 1$. This completion gives us us the algebraically closed complex Hilbert Space of quantum theory. However, no ontological significance should be given to the extended set of states so formed: by construction, they are there solely for computational convenience. Because the invariant set operators $e^{i\phi}$ are not additive for any finite $N$, they only become algebraic at the singular \cite{Berry} limit $N=\infty$. In Appendix \ref{Dirac} we show how the relativistic form of the Schr\"{o}dinger equation, the Dirac equation, can be viewed as a dynamical evolution equation in invariant set theory, in the singular limit $N=\infty$. 

The analysis provides a novel realistic interpretation of the EPR experiment (one which has nothing to do with Einstein-Rosen bridges \cite{MaldacenaSusskind}, therefore suggesting that ER$\ne$EPR). In the EPR experiment, if Alice measures the position of her particle, then the position of the second particle is determined. However, by the discussion above, it is not the case that that Alice, having actually measured position, might have measured momentum. Hence, it is not the case that the second particle must be prepared for the possibility that Alice might have measured momentum and therefore must have a well-defined momentum. 

The invariant set concept is holistic. If a point in cosmological state space does not lie on $I_U$, then nothing about the universe associated with this point is real. It doesn't matter a jot that the two particles exist on opposite sides of the universe and do not interact. Any counterfactual perturbation that leads Alice to measure momentum, takes the whole universe, including the moons of Jupiter, off $I_U$. In this sense the moons of Jupiter do care what experiment is conducted here on Earth. However, nowhere do we need to invoke a breakdown of determinism or local causality to explain this: the information about whether $U$ is a state of physical reality is encoded in the holistic but locally causal invariant set.

As discussed, the number-theoretic incommensurateness of $\phi/\pi$ and $\cos \phi$ encodes the uncertainty principle. In this sense, this account of EPR does not reveal any inconsistency with the uncertainty principle - indeed invariant set theory provides a rational explanation for it. However, because the state space of quantum theory is an algebraically complete vector space (consistent with the singular limit at $N=\infty$), quantum theory itself cannot discriminate between physically real states on the invariant set, and physically unreal states off the invariant set; its square-integrable functional form is too coarse-grain to provide such discrimination. This is essentially why quantum theory can never be considered a realistic theory. In this sense, I agree with EPR that the wave function does not and cannot provide a complete description of physical reality. 

These issues will be revisited in a more precise way when discussing the Bell Theorem in Section \ref{BellTheorem}. 

\section{Probability \emph{vs} Frequency of Outcome}
\label{measurement}

Above, we have interpreted the Hilbert vectors (\ref{momentum}) and (\ref{position}) as an uncertain selection from a sample space of neighbouring trajectories on $I_U$ in state space. However, in quantum theory the squared amplitudes of these Hilbert vectors also describe frequencies of occurrence of sequences of outcomes of similarly-prepared experiments in our unique space-time (i.e. relative to one particular history). The relationship between probability and frequency of occurrence can often be conceptually problematic in physical theory \cite{Wallace}; but not so with fractal geometry. 

\begin{figure}
\centering
\includegraphics[scale=0.5]{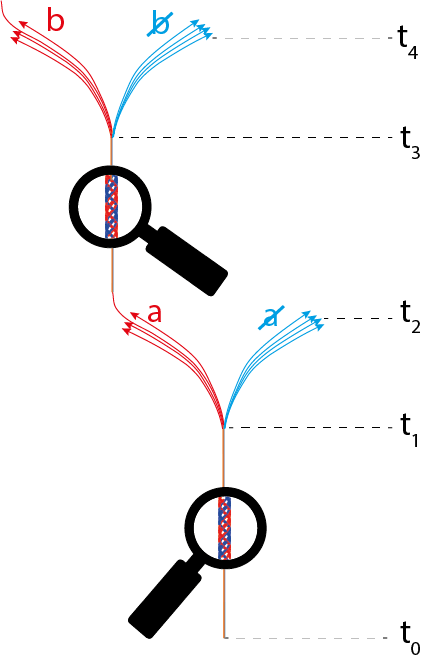}
\caption{A schematic illustration of state-space trajectories or histories (each a cosmological space-time projected into 2D) on the invariant set $I_U$, here represented locally as $C(2) \times \mathbb{R}$. Periods of chaotic evolution (between $t_1$ and $t_2$ and between $t_3$ and $t_4$) correspond to what would conventionally be referred to as periods of decoherence as the system interacts nonlinearly with its environment. Here the state-space clusters $a$, $\cancel a$, $b$ and $\cancel b$, to which the trajectories on $I_U$ are attracted, correspond to eigenstates of the relevant observable in standard quantum theory but are associated with deterministic riddled basin dynamics in invariant set theory (see Section \ref{Dynamics}). A given trajectory, under $N$-iterate magnification, reveals a neighbourhood comprising $2^N$ trajectories, consistent with the self-similar structure of $C(2)$.}
\label{magglass}
\end{figure}

To understand this, I need to outline how self-similar structure manifests itself on $I_U$. Fig \ref{magglass}, starting at  $t=t_0$ at the bottom of the Figure, shows what appears to be a single trajectory or history, i.e. an element of $C_i(2) \times \mathbb R$ for some iterate number $i$. However, looking through the magnifying glass the trajectory in fact comprises $2^N$ trajectories - shown (and reminsicent of DNA) with a compact helical structure (c.f. Appendix \ref{Dirac}). Between $t_1$ and $t_2$ these trajectories diverge chaotically into two state space clusters (described in more detail in Section \ref{Dynamics}) and the labelling above describes whether or not a trajectory evolves to cluster $a$ or cluster $\cancel a$ as a result of this chaotic evolution. The parameter $N$ describes the number of iterates of $C(2)$ that it takes to describe the clustering process between $t_1$ and $t_2$ and defines the strength of the chaotic dynamics that operates during the clustering procedure (see Sections \ref{Dynamics} and \ref{Gravity}). The physical reason for such instability can be related to the phenomenon of decoherence: on different nearby trajectories, the system interacts in different ways with its environment. As these differences grow, the trajectories diverge more from one another. In Section \ref{Dynamics}, I relate these clusters to measurement eigenstates and In Section \ref{Gravity}, I speculate that the clustering process is a manifestation of the phenomenon of gravity. 

If we take one of the trajectories of $C_{i+N}(2) \times \mathbb R$ (e.g. between $t_2$ and $t_3$), by self similarity it can also be seen to comprise a set of $2^N$ trajectories (of $C_{i+2N}(2) \times \mathbb R$) which undergoes a further period of chaotic evolution, between $t_3$ and $t_4$. And so on and so on. 

A point $X \in C(p)$ can be represented by the $p$-adic integer $...x_3x_2x_1.$ where $x_i \in \{0,1,2, \ldots p-1\}$. Here $x_i$ defines the segment of the $i$th iterate of $C(p)$ in which $X$ lies. $C(p)$ comes with a natural measure: the Haar Measure. With respect to this measure, the probability that $x_i$ equals any of the digits in $\{0,1,2, \ldots p-1\}$ is the same and equal to $1/p$. The following theorem relates probability to frequency of occurrence. 
\bigskip

$\mathbf{Theorem}$ \cite{Ruban} Let $X$ be a typical element of a Cantor set $C(p)$, i.e. an element drawn randomly with respect to the Haar measure. Then with probability one, the frequency of occurrence of any of the digits $\{0,1,2, \ldots p-1\}$ in the expansion for $X$ is equal to $1/p$. 
\bigskip

Hence, by Ruban's theorem with $p=2$, if the probability of a typical trajectory being attracted to the $a$ cluster is equal to $q/2^N$, then the frequency of occurrence of the $a$ cluster in a long sequence of similarly-prepared experiments (i.e. instability/cluster pairs) in any one trajectory is also $q/2^N$. The relationship between probability and frequency of occurrence is straightforward in a fractal setting. 

\section{The Bell Theorem}
\label{BellTheorem}

The discussion on EPR in Section \ref{interfere} is relevant to the Bell Theorem. As above, suppose Alice chooses between $x=0$ or $x=1$ and Bob $y=0$ or $y=1$. A pair of values $(x,y)$ imply a pair of spin measurements on pairs of entangled particles, where the measuring apparatuses have relative orientation $\theta_{xy}$. If $x$ and $y$ are considered as points on $\mathbb S^2$, then $\theta_{xy}$ is the angular distance between $x$ and $y$. In Appendix \ref{Hilbert}we extend the analysis in Section \ref{interfere} to show how to interpret the tensor-product Hilbert state
\be
|\psi_{ab}\rangle= \gamma_0 |a\rangle |b \rangle + \gamma_1 e^{i \chi_1} |a\rangle |\cancel{b} \rangle + \gamma_2 e^{i \chi_2}|\cancel{a}\rangle |b \rangle + \gamma_3 e^{i \chi_3}|\cancel{a}\rangle |\cancel{b} \rangle, 
\ee
where $\gamma_i, \chi_i \in \mathbb R$ and $\gamma_0^2+\gamma_1^2+\gamma_2^2+\gamma_3^2=1$, as an uncertain selection of a pair of elements $\{a_i, b_i\}$ from the bit strings
\begin{align}
S_a&=\{a_1 \; a_2 \ldots a_{2^N}\} \nonumber \\
S_b&=\{b_1 \; b_2 \ldots b_{2^N}\} 
\end{align} 
where $a_i \in \{a, \cancel a\}$, $b_i \in \{\cancel b\}$. This interpretation is only possible if $\gamma^2_i$ and $\chi_i/\pi$ are finitely describable. That is to say, the mapping from bit-string space to Hilbert Space is an injection. As discussed in Appendix \ref{Hilbert}, this implies $\cos \theta_{xy}$ must be finitely describable. 

Suppose Alice and Bob choose some particular pair $(x, y)$, so that $\{\lambda, x, y\}$ describes a universe on $I_U$. Then, based on the finite describability of $\cos \theta_{xy}$  the following can be shown: 
\begin{itemize}
\item  The counterfactual universe $\{\lambda, 1-x, 1-y\}$ also lies on $I_U$ - this history is consistent with $\{\lambda, x, y\}$ as far as invariant set theory is concerned. That is to say, if $\cos \theta_{xy}$ is finitely describable, then so too is $\cos \theta_{(1-x)(1-y)}$.
\item The counterfactual universes $\{\lambda, x, 1-y\}$ and $\{\lambda, 1-x, y\}$ (almost certainly) do not lie on $I_U$ - these histories are inconsistent with $\{\lambda, x, y\}$ as far as invariant set theory is concerned. That is to say, if $\cos \theta_{xy}$ is finitely describable, then $\cos \theta_{(1-x)y}$ and  $\cos \theta_{x(1-y)}$ are not finitely describable.
\item Let $z =x+y \mod 2$. Then, the set $\Lambda_{z=0}$ of supplementary variables $\lambda$ consistent with $z=0$ must be disjoint from the set $\Lambda_{z=1}$ of supplementary variables $\lambda$ consistent with $z=1$. 
\end{itemize}

\begin{figure}
\centering
\includegraphics[scale=0.5]{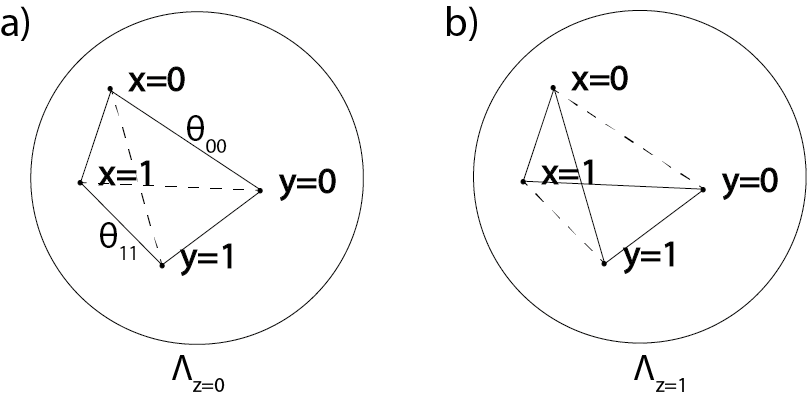}
\caption{Alice and Bob's choices $x=0$, $x=1$, $y=0$ and $y=1$ associated with measurement options for a CHSH experiment, shown schematically as four points on the 2-sphere. The lines between these points actually represent great circles where the angular distance between $x$ and $y$ is $\theta_{xy}$. Where the lines are solid (dashed), the corresponding cosines of the angular distances are (are not) finitely describable for number-theoretic reasons (see Appendix \ref{CHSH}). In a), corresponding to the sample space $\Lambda_{z=0}$ of supplementary variables, where $z=x+y \mod 2$, $\cos \theta_{00}$ and $\cos \theta_{11}$ are finitely describable. In b), corresponding to the disjoint sample space $\Lambda_{z=1}$ of supplementary variables, $\cos \theta_{01}$ and $\cos \theta_{10}$ are finitely describable. This means that the \emph{precise} positions of the $x$ and $y$ points in the left- and right-hand panels are not identical. However, to within the necessarily finite precision of the measurement orientations corresponding to the $x$ and $y$ button pushes, these points can be treated in any practical sense as if they are identical. In both figures, the cosines of the angular distances between $x=0$ and $x=1$, and between $y=0$ and $y=1$ are finitely describable because it is always possible to measure a particle pressing $x$ and then re-inject the particle into the measuring device pressing button $1-x$ (and similarly for $y$ and $1-y$). }
\label{fig:CHSH}
\end{figure}

The reasons for these conclusions combine geometry and number theory. Fig \ref{fig:CHSH} shows, separately for $\Lambda_{z=0}$ and $\Lambda_{z=1}$, the choices $x=0$, $x=1$, $y=0$ and $y=1$ represented as four points on the 2-sphere. The lines represent great circles and are solid if $\cos \theta_{xy}$ can be finitely described, and dashed otherwise. The left-hand figure corresponds to the sample $\Lambda_{z=}0$ and hence measurements where either $x=0$ and $y=0$, or $x=1$ and $y=1$. Here  $\cos \theta_{00}$ and $\cos \theta_{11}$ are finitely describable, and $\cos \theta_{01}$ and $\cos \theta_{10}$ not. The reason why the latter are not finitely describable is discussed in Appendix \ref{CHSH} and makes use of the theory of Pythagorean triples - specifically that there are no Pythagorean Triples $\{a,b,c\}$ where $c$ is a power of 2.The right-hand figure corresponds to the sample $\Lambda_{z=1}$ and hence measurements where either $x=0$ and $y=1$, or $x=1$ and $y=0$. Here  $\cos \theta_{01}$ and $\cos \theta_{10}$ are finitely describable,  and $\cos \theta_{00}$ and $\cos \theta_{11}$ not. Of conceptual importance is the implication that the \emph{precise} position of the four points cannot be absolutely identical in the left and right-hand figures. However, to within \emph{any} finite experimental precision, the position of the four points can be considered identical. Like quantum interferometry, this violation of the Bell inequality relies on finite experimental precision. As discussed in Section \ref{interfere}, these results are robust to  $p$-adic noise and hence structurally stable. 

The CHSH inequality is
\be
\label{CHSH}
|C(0,0)+C(0,1)+C(1,0)-C(1,1)| \le 2
\ee
where
\be
C(x,y)=p(a=b|x,y)-p(a \ne b|x,y)
\ee
Experimentally, each correlation is determined by a separate sub-ensemble of particles: let's say the first correlation is determined on Monday, the second on Tuesday and so on. Then, based on the results above, in any experiment which seeks to test the CHSH inequality, the supplementary variables $\lambda$ must necessarily be drawn from the two disjoint samples: $\Lambda_{z=0}$ (from which Monday and Thursday's sub-ensembles are drawn) and $\Lambda_{z=1}$ (from which Tuesday and Wednesday's sub-ensembles are drawn). 

For supplementary variables drawn from $\Lambda_{z=0}$ (left hand figure), the bit-string construction discussed in Appendix \ref{Hilbert} provides us with four bit strings, $S_0(x=0)$, $S_0(x=1)$, $S_0(y=0)$, $S_0(y=1)$, each of length $2^N$, such that
\begin{itemize}
\item Each of $S_0(x=0)$, $S_0(x=1)$, $S_0(y=0)$ and $S_0(y=1)$ has an equal number of 0s and 1s. 
\item The correlation between $S_0(x=0)$ and $S_0(y=0)$ is equal to $-\cos \theta_{00}$, by construction finitely describable.
\item The correlation between $S_0(x=1)$ and $S_0(y=1)$ is equal to $-\cos \theta_{11}$, by construction finitely describable. 
\item Because $\cos \phi_{01}$ and $\cos \phi_{10}$ are not finitely describable when $\Lambda_{z=0}$, the correlations between $S_0(x=0)$ and $S_0(y=1)$, and between $S_0(x=1)$ and $S_0(y=0)$ do not correspond to anything physically realisable on $I_U$ - they do not correspond to correlations on entangled particle measurements in an experiment to test the CHSH inequality. 
\end{itemize} 

Similarly, for supplementary variables drawn from $\Lambda_{z=1}$ (right-hand figure) we similarly have four bit strings, $S_1(x=0)$, $S_1(x=1)$, $S_1(y=0)$, $S_1(y=1)$, independent of the $S_0$ bit strings above, such that
\begin{itemize}
\item Each of $S_1(x=0)$, $S_1(x=1)$, $S_1(y=0)$, $S_1(y=1)$ has an equal number of 0s and 1s. 
\item The correlation between $S_1(x=0)$ and $S_1(y=1)$ is equal to $-\cos \theta_{01}$ by construction finitely describable.
\item The correlation between $S_1(x=1)$ and $S_1(y=0)$ is equal to $-\cos \theta_{10}$, by construction finitely describable. 
\item Because $\cos \phi_{00}$ and $\cos \phi_{11}$ are not finitely describable when $\Lambda_{z=1}$, the correlations between $S_1(x=0)$ and $S_0(y=0)$, and between $S_1(x=1)$ and $S_0(y=1)$ do not correspond to anything physically realisable on $I_U$ - they do not correspond to correlations on entangled particle measurements in an experiment to test the CHSH inequality. 
\end{itemize}

Since $C(x,y)=-\cos \theta_{xy}$ on $I_U$, invariant set theory violates the Bell inequality as does quantum theory. Formally, it can do this because of superdeterminism: $\Lambda_{z=0}$ is disjoint from $\Lambda_{z=1}$. However, this is not \emph{ad hoc} superdeterminism, but superdeterminism emergent from our number-theoretic/fractal geometric approach to consistent histories. This raises an important point. The need to mention here (and in Section \ref{interfere}) the reality or otherwise of certain counterfactual worlds may seem puzzling to some. After all, modern accounts of Bell's theorem (e.g. \cite{Brunner}) make no mention at all about the notion of counterfactual reality - as such the notion may seem an irrelevance. Why, then, does counterfactuality seem to play such a crucial role in the invariant set model's evasion of Bell inequalities? The answer is that, as mentioned, in the invariant set model, superdeterminism does not arise by fiat, it is a consequence of deeper number-theoretic/fractal geometric principles. Presenting (\ref{mi}) as axiomatic (or, at least, not acknowledging that it might arise, or fail, from something deeper) is again prejudicial, because it immediately suggests that any failure of (\ref{mi}) is \emph{ad hoc}. Although modern accounts of Bell's theorem are very compact, I believe that this compactness is a hindrance to the emergence of physically plausible alternatives to quantum theory. 

The transcendental property of the cosine function is essential if invariant set theory is to violate the CHSH inequality. For example, suppose the invariant set's correlations were given instead by some $F(\theta)$, where $F$ was a polynomial with finitely describable coefficients. Then if both $F(\theta_1)$ and $F(\theta_2)$ were finitely describable, so too would be $F(\theta_1+\theta_2)$. In this case, the argument for disjoint sample spaces $\Lambda_{z=0}$ and $\Lambda_{z=1}$ would fail. In turn, the model would have to be constrained by the CHSH inequality (i.e. be essentially classical) and would therefore be inconsistent with experiment. With such polynomial functions, it would also be possible to create sub-systems, and the essential holism of the model would also fail. 

It is well known that it is possible to concoct `superquantum' theories where the CHSH inequality is violated more that does quantum theory \cite{Popescu,Tsirelson}. Could one concoct a type of invariant set theory in which the CHSH inequalities are maximally violated? The answer depends on whether there exist other transcendental functions $T$ (in addition to the cosine function) such that if $T(\theta_1)$ and $T(\theta_2)$ are finitely describable, then $T(\theta_1+\theta_2)$ is almost certainly not, and where $T(0)=1$, $T(\pi/2)=0$, $T(\pi)=-1$. The failure to find an alternative suitable transcendental correlation function would provide a new approach to understanding the Tsirelson bound.  

\section{Random Bits and Free Will}
\label{free}

As discussed, if $U$ is a universe on $I_U$ where $x=0$ and $y=0$, then the counterfactual universe $U'$ with the same supplementary variables $\lambda$, but where $x=1$ and $y=0$, does not lie on $I_U$. Let us suppose that $x$ is set by Bell's pseudo-random number generator, i.e. the value of the millionth bit in $U$ is a 0, and the value of the millionth bit in $U'$ is a 1. If $U$ lies on $I_U$ at the time $x$ is set, then it also lies on $I_U$ at the earlier time the million-digit variable is input to the pseudo random number generator. This is true, not because of any notion of retrocausality \cite{Price:1997}, because the notion of dynamical invariance is necessarily independent of time: if a state lies on the invariant set now, it \emph{always} has lain on it, and \emph{always} will lie on it. Similarly, if a state does not lie on $I_U$ now it \emph{never} has lain on it and \emph{never} will lie on it. In particular $U'$ did not lie on $I_U$ when the million-digit variable with its last counterfactually-flipped bit was input to the pseudo-random number generator. 

The reason, then, why Bell's argument about physical randomisers may be incorrect is that in this invariant set theoretic approach, the difference between $U$ and $U'$ at the time the million-digit variable was input to the pseudo-random number generator is not (2-adically) small. Indeed, reducing the bit which sets $x$ from a millionth to a trillionth will not affect the  2-adic distance of $U'$ from $I_U$ one jot (even though it makes the Euclidean distance smaller). Bell \cite{Bellb} recognised that his arguments were not water tight, commenting:
\begin{quote}
Of course it might be that these reasonable ideas about physical randomisers are just wrong- for the purposes at hand. A theory may appear in which such conspiracies inevitably occur, and these conspiracies may then seem more digestible than the non-localities of other theories. When that theory is announced, I will not refuse to listen, either on methodological or other grounds.
\end{quote}
As discussed above, there are no conspiracies in invariant set theory: there is nothing secretive about the invariant set, even though many of its properties may be non-computable. 

In contemporary attempts to test (\ref{mi}), $x$ and $y$ are often set by for-all-practical-purposes random bits rather than direct experimenter choices e.g. \cite{Shalm} extracted bits from the movie \emph{Back to the Future}. Exactly the same arguments as used above apply to these movie bits. Hence, if a certain bit of the movie determines $x$, then flipping the bit keeping the supplementary variables $\lambda$ fixed, takes the whole state of the universe off the invariant set to a non-ontological state of physical unreality. If $\{\lambda, x=0\}$ denotes a state of the universe on $I_U$, where the relevant movie bit from \emph{Back to the Future} is a 0, then $\{\lambda, x=1\}$ denotes the state $U'$ which does not lie on $I_U$. As before, consistent with the holism of the theory, none of the components of $U'$, such as the moons of Jupiter, have physical reality. Their reality (and that of all the clusters of galaxies in distant parts of the universe) can be obliterated at an instant by just counterfactually flipping that one movie bit - as discussed, $p$-adically this bit flip is not a small-amplitude perturbation. The fact that these moons and clusters are spacelike separated from the place where the counterfactual bit flip occurs is a complete irrelevance because the counterfactual bit flip violates a global state-space constraint (that states of reality lie on a global invariant set). This discussion can be thought of as an illustration of the Takens embedding theorem in dynamical systems theory \cite{Takens:1981}: the whole invariant set can be constructed from a sufficiently long time series of just one component of the invariant set, even if that component is spatially localised and energetically insignificant. 

Suppose at the last minute, the experimenters decide that $x$ and $y$ will be determined by bits from \emph{Jaws} rather than \emph{Back to the Future}. Can we still say that counterfactually flipping the bit from \emph{Back to the Future} will cause the state of the universe to move off the invariant set? No we can't, because now the bit from \emph{Back to the Future} is no longer the determinant to a future experiment involving non-commutative operators. Hence, we can no longer invoke the deterministic invariance argument that because counterfactually flipping $x=0$ to $x=1$ takes the universe off the invariant set, counterfactually flipping the \emph{Back to the Future} bit must necessarily do the same. 

As discussed, Bell used randomisers as a way to avoid discussing the contentious metaphysics of human free will. However, fundamentally, there are no new issues to discuss if we replace these randomisers with human brains. Human cognition can be just as susceptible to pseudo-random (but ultimately deterministic) processes inside the brain, as does Bell's pseudo-random number generator. These arise not least because of the extreme slenderness of human axons, making the protein transistors which amplify electric signals propagating along these axons, subject to thermal noise \cite{RollsDeco}. Indeed this susceptibility, born out of a need for the brain with its hundred-billion neurons to be outstandingly energy efficient (it operates with the power needed to light a domestic light bulb), could be what makes us creative \cite{PalmerOShea}. If the decision to measure $x=0$ rather than $x=1$ depends sensitively on the action of one of these protein transistors, then the consequences for lying on the invariant set in the counterfactual world where a particular ion failed to activate a particular protein transistor, are no different to the pseudo-random number bits and movie bits that have been discussed above. As has been argued by such eminent philosophers as Thomas Hobbes, David Hulme and John Stuart Mill \cite{Kane}, our sense of free will merely reflects an absence of constraints preventing us from doing what we want to do. From this point of view, neither determinism nor indeed superdeterminism is an impediment to free will. Positing superdeterminism as a constraint on free will is simplistic and unnecessarily restrictive on the class of superdeterministic theories. 

In Appendix \ref{PBR}, I discuss the PBR Theorem \cite{Pusey} from this superdeterministic perspective (where violation of Measurement Independence is traded for violation of Preparation Independence). 

I do not think it will ever be possible to test the Measurement Independence assumption directly in these Bell test experiments. We have to find other ways. These may involve experimental studies, or direct cosmological observations, where the effects of neither quantum nor gravitational physics are negligible. We briefly address such matters below.   

\section{Fooled by Our Genes?}
\label{Gene}

Let me address a question I have been asked by one eminent quantum foundations expert. Why would nature be so incredibly devious to lead us, a seemingly intelligent species capable of some astonishing achievements in both the arts and the sciences, to be so comprehensively fooled into thinking that the world around us is not locally real, when it actually is? 

Actually I do not believe nature is the least bit devious. Rather, I believe we have been fooled by our genes. Consider our first experiences as human beings. A baby in the cot sees a colourful toy. For one or more reasons, she is genetically programmed to be attracted to such colourful objects, and so instinctively wants to explore the toy further. To do this, she has to get a part of her body (typically mouth and/or hands) in close proximity to the toy. In so doing, she implicitly learns a fundamental fact about the nature of physical space: closeness is synonymous with smallness of Euclidean distance. If this sense of spatial awareness is the first thing we learn as humans, it may be the hardest thing to let go of, when, in later life, we come to explore more abstract spaces, notably state spaces. Hence, when the philosopher David Lewis says that one world is closer to actuality than another if the first resembles the actual world more than the second does (c.f. Section \ref{metric}), I believe Lewis is inadvertently misapplying intuitive ideas learned in the cot. 

Ironic then that a theoretically sound class of theory has been rejected, supposedly because of concerns about subversion from effects worse than alien mind control, but in fact because our minds have been comprehensively subverted by something much closer to home - gene control! 

\section{Invariant Set Theory \emph{vs} Cellular Automaton Theory}
\label{tHooft}

It was mentioned in Section \ref{Dislike} that there is a notable exception to the community's rejection of superdeterminism \cite{tHooft:2015b, tHooft:2015}. 't Hooft argues that it is possible to formulate a classical model of the real world which underpins quantum physics. He proposes a realistic deterministic cellular automaton dynamical system in place of the Schršdinger equation. This system is presumed to describe the evolution of the world with respect to in a unique `ontological basis'. The types of  counterfactual worlds discussed above are deemed unrealistic simply because, by fiat, the corresponding basis is not ontological. 

There is a fundamental theoretical difference between cellular automaton theory and invariant set theory. As a model of quantum physics, cellular automaton theory requires two separate elements: a deterministic (Schr\"{o}dinger-like) dynamical evolution equation, $D$, and a constraint on possible initial conditions $I$.  In classical theories (which includes 'tHooft's theory), the dynamical laws of evolution and the initial conditions from which future states evolve, are largely independent of one another. For example, the classical Lorenz equations \cite{Lorenz:1963} can be integrated from any point in the Euclidean state space $\mathbb R^3$ of the model. In classical theory, we can constrain the initial state to some region of state space, e.g. $X \ge 0$, but such a constraint must be applied independently of, and in addition to, the dynamical equations themselves. To constrain initial conditions 't Hooft considers the special nature of the early universe \cite{tHooft:2015}:
\begin{quote}

\ldots even in a superdeterministic world, contradictions with Bell's theorem would ensue if it would be legal to consider a change of one or a few bits in the beables describing Alice's world, without making any modifications in Bob's world. \ldots [However,] it is easy to observe that, certainly in the distant past, the effects of such a modification would be enormous and it may never be compatible with a simple low-entropy Big Bang \ldots Thus, we can demand in our theory that a modification of just a few beables in Alice's world without any changes in Bob's world is fundamentally illegal. This is how an ontological deterministic model can `conspire' to violate Bell's theorem. 

\end{quote}

There are similarities with the arguments above. However, notwithstanding the fact that the reasons for the low-entropy of the early universe remain controversial \cite{Penrose:2010}, 'tHooft's model raises a conceptual problem: Why has nature chosen to constrain both $D$ to be Schr\"{o}dinger like, \emph{and} at the same time, only permitted a single initial state, the ontological initial state? Of course without the constraint on $I$, the Bell inequalities would not be violated, and without the constraint on $D$, we would observe basic quantum phenomena like black body radiation. Hence, both these constraints are necessary to account the world around us. But why both? There seems to be something unnecessarily complex and theoretically perplexing that we have to invoke two seemingly separate and independent constraints to arrive at a description of the observed world. 

In invariant theory, $D$ and $I$ are not independent. They are both subservient to the underpinning fractal geometry of the invariant set. That is to say, consistent with invariant set theory not being a classical theory, there aren't two separate constraints, but only one: the geometry of $I_U$. For this reason, invariant set theory is conceptually a simpler theory than cellular automaton theory. Indeed, rather than relying on the low entropy of the early universe, as discussed in Section \ref{Gravity}, it is possible that the low entropy of the early universe may in fact be derivable from properties of the invariant set.

\section {Riddled Basins, The Schr\"{o}dinger Equation as a Singular Limit, and Why Stochasticity Fails.}
\label{Dynamics}

\begin{figure}
\centering
\includegraphics[scale=0.4]{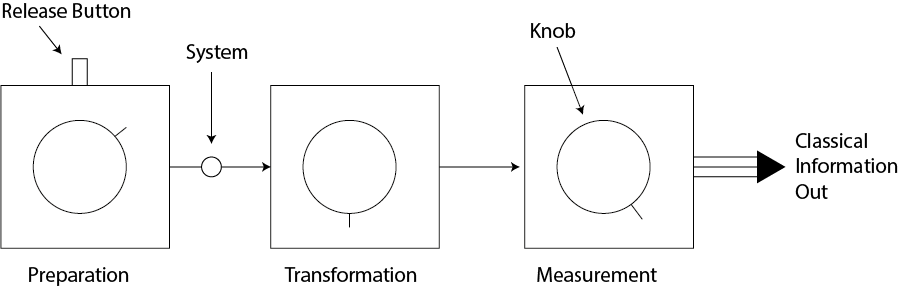}
\caption{A schematic of a typical situation where a system is prepared in some initial state and subject to a transformation and measurement. From \cite{Hardy:2004}}
\label{Hardy}
\end{figure}

The discussion so far has mostly been about the kinematics of the invariant set. How do we describe dynamical evolution? To frame the problem, let us use the canonical experimental situation where a system is prepared by a device with a knob for varying the state of the system produced and a release button for releasing the system, a transformation device for transforming the state (and a knob to vary the transformation), and a measuring apparatus for measuring the state (with a knob to vary what is measured) which outputs classical information. See Fig \ref{Hardy} (from \cite{Hardy:2004}). 

In invariant set theory, this experimental situation is described graphically by one of the trajectories in Fig \ref{magglass} which, let's say, belongs to the $a$ cluster at time $t_2$ (i.e. quantum mechanically, is prepared in the $|a\rangle$ state at $t_2$) and belongs to either the $b$ cluster or the $\cancel b$ cluster at time $t_3$ (i.e. quantum mechanically, reduces to the $|b\rangle$ or $| \cancel b \rangle$ state at $t_3$). This description is consistent with the notion that in invariant set theory the most primitive expression of the laws of physics is a description of the geometry of the invariant set $I_U$ in state space. 

In invariant set theory, dynamical evolution associated with divergence and clustering are associated with deterministic processes. It is known that unpredictable evolution towards multiple attractors can be described deterministically by so-called riddled-basin dynamics \cite{Alexander:1992}. This arises when a chaotic oscillator is coupled to a model with multiple potential wells. With such coupling, the basin boundaries can become fractally intertwined and it is generally unpredictable whether the system lies in one basin of attraction or another other. Chaotic evolution can be based on nonlinear $p$-adic mappings, much like the logistic maps on $\mathbb R$ \cite{WoodcockSmart}. Here is enough to utilise a simple Bernoulli or binary shift map - a type of logistic map -  on the 2-adic bits representing a state on $I_U$. This shift map would generate the divergence of trajectories shown in Fig \ref{magglass}, and the grouping into the two clusters $b$ and $\cancel b$ is then consistent with two riddled basins of attraction on the invariant set. The parameter $N$ describes the number of iterations of the binary shift needed to lead to evolution to a cluster. With $N$ large, the fate of an individual trajectory at $t_2$ is unpredictable. This evolution is reminiscent of the notion of `objective reduction' in quantum physics. However, such a description would be misleading because invariant set theory does not assume any fundamental ontological significance to the superposed Hilbert state, and hence, as such, no `reduction' is actually taking place. Rather, the riddled-basin procedure combines a quasi-linear instability process, which separates nearby trajectories on $I_U$ due to way in which neighbouring histories interact with their environment (i.e. decohere) and with a nonlinear clustering process. In Section \ref{Gravity} I speculate that the clustering (or 'clumping') of trajectories in state space is a manifestation of the phenomenon we call gravity. 

In the literature, objective representations of the measurement procedure have been described by stochastic mathematics \cite{Pearle:1976, Ghirardi, Percival:1995}. However, in this framework, conventional stochastic dynamics would completely destroy the properties of the invariant set which allow it to violate the Bell Theorem in a locally causal context.  This may at first sight seem surprising given that stochasticity makes no difference to the interpretation of the Bell inequalities for conventional hidden-variable models, if the dynamics of these models is deterministic or stochastic \cite{Bell}. However, here, if the measure-zero nature of the invariant set is stochastically smeared out onto the full measure of the Euclidean space in which $I_U$ is embedded, then the model will no longer have the property that a small (in the Euclidean sense) perturbation can take a point on the invariant set off it. 

The preparation procedure can be viewed in a similar way.  As a result of the preparation procedure, the system, at the time of preparation, lies in one of the state-space clusters associated with preparation ($a$ and $\cancel a$). When the system is released, it emerges from the preparation procedure with (say) the property `I have originated from cluster $a$'.

In quantum theory, the transformation from $|a\rangle$ to either $|b\rangle$ or $|\cancel b \rangle$ is given by the Schr\"{o}dinger equation. In Appendix \ref{Dirac} we show how the simplest form of the Dirac equation arises in invariant set theory in the singular limit \cite{Berry} $N= \infty$. For finite $N$, time evolution of the bit strings $S$ are effected by the cyclical permutation operator $\zeta$ described in (\ref{zeta}). Because $\zeta^{2^N} S_b= S_b$, this evolution is inherently oscillatory in time. A key physical relationship associated with the Schr\"{o}dinger equation is between the frequency of this oscillation and the mass and hence energy of the associated particle: $E=\hbar \omega$. With energy having an expression in terms of space-time curvature through the field equations of general relativity, this can be seen as an example of a fundamental relationship
\be
\label{generalisedequation}
\parbox{2.5in}{Expression for a property of the locally Euclidean space-time geometry $\mathcal M_U$}
\ \ 
=
\ \ \ 
\parbox{2.5in}{Expression for a property of the locally $p$-adic state-space geometry $I_U$ }
\ee
with Planck's constant providing the basic constant of nature that translates properties of state-space geometry into properties of space-time geometry. We discuss further possible implications of this relationship in Section \ref{Gravity}. 

In quantum theory, the Schr\"{o}dinger equation provides the dynamical underpinning for the transformation phase of Fig \ref{Hardy}. The equation has of course proven exceptionally accurate. Hence, rather than add extra terms to this equation, invariant set theory provides restrictions on the equation by not allowing the full Hilbert Space of states but only an algebraically open subset (where complex phases are finitely describable multiples of $\pi$ and where space-time orientations have finitely describable cosines). This restriction does not destroy the Schr\"{o}dinger equation's key property of describing conservation of probability (just as the classical Liouville equation still describes conservation of probability when state-space trajectories are restricted to a fractal attractor). The precise linearity of the Schr\"{o}dinger equation reflects this conservation of probability, and this linearity is not in any way perturbed by the presence of the highly nonlinear riddled-basin dynamics discussed above (just as the linearity of the Liouville equation in classical physics is not challenged by underpinning nonlinear dynamical evolution). 

This suggests a new interpretation of the Schr\"{o}dinger equation: as a computationally powerful tool, but one whose functional analytic form is too coarse grained to be able to distinguish ontological and non-ontological states, and therefore too coarse grained to describe quantum ontology. Partial analogies arise in classical physics where it is computationally convenient to represent properties of some fundamentally discrete system, e.g. molecules of a fluid, using variables drawn from the reals, e.g. the fluid variables of the Navier-Stokes equation. However, there is a crucial difference between this classical situation and the one proposed here. In classical physics we use the Euclidean metric in both space time and state space. Hence the continuum approximation is arbitrarily good if the mean free path of molecules between collisions is small enough. However, in invariant set theory, with its use of the $p$-adic metric as a yardstick of distance in state space, this is no longer true. In the latter framework at the ontological level, it does matter whether we are dealing with the original discrete variables or the more computationally convenient continuum variables, no matter how large the (finite) parameter $N$. In particular, a realistic theory based on continuum variables must necessarily be nonlocal in order to violate the Bell inequalities. 

In this sense, one can interpret the quantum potential of Bohmian theory as a coarse-grained $L^2$ representation of the invariant set in configuration space. Again, with such a realistic representation, the Bohmian quantum potential does not have the fine-grained ontological properties of the invariant set and therefore Bohmian theory has to be explicitly nonlocal. 

\section{Quantum Gravity and the Dark Universe - The End of Particle Physics?}
\label{Gravity}

The primary motivation for developing the model above was based on a growing belief that the fundamental impediment to synthesising quantum and gravitational physics is in fact quantum theory \cite{Penrose:1976}. If a causal geometric model of quantum physics can be developed, it may, superdeterministic or not, stand a much greater chance of being synthesised with Einstein's causal geometric model of gravity, than does quantum theory. Here I wish to suggest some possible implications of invariant set theory for such a synthesis. The remarks below are, of course, very speculative. 

As discussed, the geometric basis of the proposed superdeterministic model is of a fractal invariant set in the state space of the universe. Such invariant sets (e.g. the Lorenz attractor), comprise just one indefinitely long trajectory wrapped up in a compact set. By comparison then, $I_U$ comprises the trajectory of a mono-universe (i.e. not a multiverse) evolving over multiple epochs on a similar compact set. From this perspective, the sample space of neighbouring trajectories over which probabilities are defined, merely define instances of our universe at earlier or later epochs. There are no `Many Worlds' as in the Everett interpretation - just one. Whether the universe is open or closed is currently an unresolved question; it appears to be on the borderline. The theory proposed here suggests an underlying finite and hence closed universe. 

Continuing the notion of finiteness, although the discussion so far has assumed that $I_U$ is  precisely fractal, there is nothing in the theory that would prevent $I_U$ from being a sufficiently complex finite limit cycle (that approximates well a fractal). That is to say, it could be that the self-similarity discussed above, only persists to some large but finite number of iterations. 

By treating the geometry of $I_U$ as a primitive expression of the laws of physics, then, as suggested by (\ref{generalisedequation}), we can expect that the pseudo Riemannian geometry of our space time is influenced by the fractal geometry of $I_U$. That is to say, the geometry of space time should be partially influenced by the geometry of neighbouring state-space space-times (i.e. our universe at later or earlier epochs).  Could this provide an explanation of dark matter in our space time?  That is to say, could it be that what we call dark matter in our space-time, is merely ordinary matter on neighbouring space-time trajectories on $I_U$, whose influence on our space time arises from the particular geometric form of $I_U$?

Indeed, following (\ref{generalisedequation}), perhaps the curvature of space-time is intimately linked to the clustering of trajectories  on $I_U$ as shown in Fig \ref{magglass}. This would suggest, consistent with earlier speculations by Penrose \cite{Penrose:1989} and Di\'{o}si \cite{Diosi:1989}), that gravity should itself be an intrinsically decoherent process. An experimental confirmation of this would provide a strong indication that the conventional quantum field theory approach to quantum gravity is misguided. Now this clustering is certainly a nonlinear process, and therefore would not occur if differences in space-time energy were sufficiently small. This idea can be made quantitative. If $E_G$ denote the gravitational self energy of the difference between mass/energy distributions in two space-times \cite{Penrose:2004}, then these space-times can be considered gravitationally indistinct (over a time interval $\tau$) if
\be
\int_{\tau} E_G dt < \hbar
\ee
Two space-times that are sufficiently similar would then be gravitationally indistinct. One can speculate that two space-times which differ merely by vacuum fluctuations would be gravitationally indistinct in this sense. This would imply that vacuum fluctuations do not couple to gravity and this can help explain why the cosmological constant is not 120 orders of magnitude bigger than it is. The value that the cosmological constant does take may be the cosmological consequence of (\ref{generalisedequation}), together with the generic and ubiquitous divergence of trajectories on $I_U$ (shown in Fig \ref{magglass}). This provides a new proposal for the origin of dark energy in $\mathcal M_U$. 
 
If the ideas above are correct, there will be no quantum field theoretic excitation (i.e. `particle') associated with either dark matter or dark energy. Indeed there will additionally be no such particle as a graviton. Invariant set theory, if correct, implies a limit to the ability of particle physics to explain everything we see in the world around us and indeed may signal the end of particle physics beyond the Standard Model.  However, the Standard Model has been extraordinarily successful in explaining non-gravitational physics. What role does that model have in invariant set theory? I would argue that the gauge groups of the Standard Model define the state space in which $I_U$ is embedded. From this perspective it would be conceptually wrong to imagine gravity as some extension of the Standard Model's gauge-group structure, e.g. as some kind of Yang-Mills theory. Rather we should think of gravity as the geometric phenomenon of clustering or clumping of histories on the non-trivial fractal measure-zero geometry in the space spanned by the degrees of freedom of the Standard Model's gauge group. 

One of the reasons for developing a so-called quantum theory of gravity is that it will eliminate the occurrence of space-time singularities. Equation (\ref{generalisedequation}), whereby neighbouring state-space trajectories influence our space-time, could also eliminate singularities by smearing out what would otherwise be a delta function in curvature in our space time (corresponding to the singular limit at $N=\infty$), into a finite gaussian-like function (at finite but large $N$) with support on some neighbourhood of histories in state space.

The notion of a fractal invariant set has many interesting implications for the perplexing question of time asymmetry and the Second Law of Thermodynamics. The growth of entropy from the early phase of a cosmological epoch would, as in standard chaotic dynamics, be associated with the generic divergence of trajectories shown in Fig \ref{magglass}. What causes the low entropy at the beginning of a cosmological epoch? In invariant set theory, it must be associated with state-space convergence of trajectories in the late phase of the previous epoch (and hence nothing to do with inflation). In conventional nonlinear dynamics, such convergence is typically associated with dissipation. However, here we would need to find something more fundamental. Penrose \cite{Penrose:2010} has speculated that such state-space convergence would generically be associated with the formation of black holes - more generally of collapse towards a final-time singularity (a quasi-singularity from the paragraph above).  It is this process of state-space convergence that gives rise to the lacunae which allow invariant set theory to violate Bell inequalities in the laboratory today.

These ideas in turn have implications and new perspectives on the problem of black-hole information loss: in invariant set theory information is not lost, but is compressed to the extent that the information, like vacuum fluctuations, no longer has a distinct gravitational signature. The information reacquires its gravitational signature during the divergence phase when entropy is again increasing. 

As mentioned, these remarks are speculative. However, they illustrate the fact that the development of a superdeterminstic theory of quantum physics may open up some very new ways of thinking about some of the deepest problems of contemporary fundamental physics.

\section{Conclusions}

It is time for superdeterminism and the corresponding Measurement Independence assumption to be reappraised. Visceral arguments  should be replaced with more logical and reasoned arguments. A key part of this reappraisal is the role the $p$-adic metric plays as the yardstick of choice in state space. The implications are potentially enormous, not only for our understanding of quantum physics, but even more for resolving many of the conceptual difficulties that seem to be creating a crisis of understanding in contemporary foundational physics.  

Following the Shalm et al experiments, media headlines proclaimed that the Einstein/Bohr debate has finally been settled in favour of Bohr. The theory described in this paper provides a meeting ground for both Einstein and Bohr's views and suggests that the the result of the debate may in fact be a dead heat. 

\bigskip

\bibliography{mybibliography}

\appendix
\section{$p$-adic Integers and Cantor Sets}
\label{padic}

By way of introduction to the $p$-adic numbers, consider the sequence
\be
\{1, 1.4, 1.41., 1.414, 1.4142, 1.41421 \ldots\} \nonumber
\ee
where each number is an increasingly accurate rational approximation to $\sqrt 2$. As is well known, this is a Cauchy sequence relative to the Euclidean metric $d(a,b)=|a-b|$, $a$, $b \in \mathbb{Q}$. 

Surprisingly perhaps, the sequence
\be
\{1, 1+2, 1+2+2^2, 1+2+2^2+2^3, 1+2+2^2+2^3+2^4, \ldots\}
\ee
is also a Cauchy sequence, but  with respect to the ($p=2$) $p$-adic metric $d_p(a,b)=|a-b|_p$ where
\be
|x|_p=\left \{%
\begin{array} {ll}
p^{-\textrm{ord}_p x} &\textrm{if } x \ne 0 \\
0 &\textrm{if } x=0
\end{array}%
\right.
\ee
and
\be
\textrm{ord}_p x= \left \{%
\begin{array}{ll}
\textrm{the highest power of \emph p which divides \emph x, if } x \in \mathbb Z \\
\textrm{ord}_p a - \textrm{ord}_p b, \textrm{ if } x=a/b, \ \ a,b \in \mathbb Z, \ b \ne 0
\end{array}%
\right.
\ee
Hence, for example
\be
d_2(1+2+2^2, 1+2)=1/4,\ \ d_2(1+2+2^2+2^3, 1+2+2^2)=1/8
\ee
Just at $\mathbb{R}$ represents the completion of $\mathbb{Q}$ with respect to the Euclidean metric, so the $p$-adic numbers $\mathbb{Q}_p$ represent the completion of $\mathbb{Q}$ with respect to the $p$-adic metric. A general $p$-adic number can be written in the form
\be
\sum_{k=-m}^{\infty} a_k p^k
\ee
where $a_{-m} \ne 0$ and $a_k \in \{0,1,2, \ldots, p-1\}$. The so-called $p$-adic integers $\mathbb Z_p$ are those $p$-adic numbers where $m=0$.  

It is hard to sense any physical significance to $\mathbb Z_p$ and the $p$-adic metric from the definition above. However, they acquire relevance in invariant set theory by virtue of their association with fractal geometry. In particular, the map $F_2: \mathbb Z_2 \rightarrow C(2)$ 
\be
F_2: \sum_{k=0}^{\infty} a_k2^k \mapsto \sum_{k=0}^{\infty} \frac{2a_k}{3^{k+1}} \textrm{ where } a_k \in \{0,1\}
\ee  
is a homeomorphism \cite{Robert}, implying that every point of the Cantor ternary set can be represented by a 2-adic integer. More generally, 
\be
F_p: \sum_{k=0}^{\infty} a_kp^k \mapsto \sum_{k=0}^{\infty} \frac{2a_k}{(2p-1)^{k+1}} \textrm{ where } a_k \in \{0,1, \ldots p-1\}
\ee
is a homeomorphism between $\mathbb Z_p$ and $C(p)$ To understand the significance of the $p$-adic metric, consider two points $a, b \in C(p)$. Because $F_p$ is a homeomorphism, then as $d(a,b) \rightarrow 0$, so too does $d_p(\bar a, \bar b)$ where $F(\bar a)=a, \ F(\bar b)=b$. On the other hand, suppose $a \in C(p)$, $b \notin C(p)$. By definition, if $b \notin C(p)$, then $\bar b \notin \mathbb Z_p$. Let us assume that $b \in \mathbb Q$ (see Footnote 3 above). Then $\bar b \in \mathbb Q_p$. This implies that $d_p(\bar a,\bar b) \ge p$. Hence,  $d(a,b) \ll 1 \centernot \implies d_p(\bar a, \bar b) \ll 1$. In particular, it is possible that $d_p(\bar a, \bar b) \gg 0$, even if $d(a,b) \ll 0$. From a physical point of view, a perturbation which seems insignificantly small with respect to the (intuitively appealing) Euclidean metric, may be unrealistically large with respect to the $p$-adic metric, if the perturbation takes a point on $C(p)$ and perturbs it off $C(p)$. The $p$-adic metric recognises the primal ontological property of lying on the invariant set. The Euclidean metric, by contrast, does not. 

Let $g(x, x')$ denote the pseudo-Riemannian metric on space-time, where $x, x' \in \mathcal M_U$. By contrast let $g_p(\mathcal M_U, \mathcal M_U')$ denote a corresponding metric in $U$'s state space, transverse to the state-space trajectories. As above, we suppose that if $\mathcal M_U \in I_U$, then $g_p(\mathcal M_U, \mathcal M_U') \rightarrow 0$ only in the $p$-adic sense, ie. only if $\mathcal M_U' \in I_U$. 

\section{A Key Number-Theoretic Property of the Cosine Function}
\label{number}
A key number theorem for this paper is this:
\\
\\
$\mathbf{Theorem}$\cite{Niven, Jahnel:2005}.  Let $\phi/\pi \in \mathbb{Q}$. Then $\cos \phi \notin \mathbb{Q}$ except when $\cos \phi =0, \pm 1/2, \pm 1$. 
\\
\\
$\mathbf{Proof}$. Assume that $2\cos \phi = a/b$ where $a, b \in \mathbb{Z}, b \ne 0$ have no common factors.  Since 
\be
\label{cosidentity}
2\cos 2\phi = (2 \cos \phi)^2-2
\ee
then 
\be
2\cos 2\phi = \frac{a^2-2b^2}{b^2}
\ee
Now $a^2-2b^2$ and $b^2$ have no common factors, since if $p$ were a prime number dividing both, then $p|b^2 \implies p|b$ and $p|(a^2-2b^2) \implies p|a$, a contradiction. Hence if $b \ne \pm1$, then the denominators in $2 \cos \phi, 2 \cos 2\phi, 2 \cos 4\phi, 2 \cos 8\phi \dots$ get bigger without limit. On the other hand, if $\phi/\pi=m/n$ where $m, n \in \mathbb{Z}$ have no common factors, then the sequence $(2\cos 2^k \phi)_{k \in \mathbb{N}}$ admits at most $n$ values. Hence we have a contradiction. Hence $b=\pm 1$ and $\cos \phi =0, \pm1/2, \pm1$. 
\\
\\
If, moreover, $\phi/\pi$ is describable by a finite number of bits, then  $\cos \phi =0, \pm 1$

\section{A Realistic Interpretation of Complex Hilbert Vectors}
\label{Hilbert}
\subsection{One Qubit}
Extending the discussion in Section \ref{interfere}, we view the Hilbert vector
\be
\label{1qubit1}
|\psi_{a}(\theta, \phi)\rangle=
\cos \frac{\theta}{2}|a\rangle
+e^{i\phi} \sin \frac{\theta}{2} |\cancel{a}\rangle
\ee
as an uncertain element $a_i \in \{a, \cancel a\}$ selected from the bit string $S_a(\theta, \phi)=\{a_1, a_2, \ldots a_{M}\}$.  By this is meant that there is a deterministic procedure which selects an element of $S_a(\theta, \phi)$ but this  procedure is sufficiently unpredictable as to be unknowable (Section \ref{Dynamics}). If the qubit is associated with a pure state, then $M=2^N$ where $N$ is related to the number of fractal iterates needed to evolve to a state-space cluster. The probability of selecting any particular element $a_i$ is independent of $i$. As discussed in the main body of the text, the additive form of (\ref{1qubit1}) results from Pythagoras' theorem. 

Let us start with $\phi=0$. Then  
\be
S_a(\theta, 0)=\{\underbrace{a\;a\;a\ldots a}_{M\cos^2\theta/2}\ \underbrace{\cancel a\;\cancel a\;\cancel a\ldots \cancel a}_{M\sin^2\theta/2}\}
\ee
where $\cos^2 \theta/2$ is a rational number of the form $M'/M$ for integer $0 \le M'\le M$. For a pure state, then in addition to being rational, $\cos^2 \theta/2$ and hence $\cos \theta$ must be finitely describable.

To define $S_a(\theta,\phi)$ where $\phi \ne 0$, let
\be
\label{zetan}
\zeta \{a_1\; a_2\; \ldots a_{M}\}=\{a_M \;a_1 \; a_2\; \ldots a_{M-1}\}.
\ee
Since $\zeta^M  \{a_1\; a_2\; \ldots a_{M}\}=  \{a_1\; a_2\; \ldots a_{M}\}$ we can treat $\zeta$ as an operator representation of an $M$th root of unity and write $\zeta=e^{2\pi i /M}$. Using this (\ref{1qubit1}) is defined as an uncertain element of
\be
S(\theta, \phi)=\zeta^{M''}S_a(\theta, 0)
\ee
for integer $M''$, where $\phi=2\pi M''/M$. When $M=2^N$ then $\phi/\pi$ must be finitely describable. Since the selection procedure is assumed chaotic, the uncertain element selected will be sensitive to the ordering of bit-string elements. Hence, for example, the chaotic procedure will not in general select the same element when applied to the bit string $S_a(\theta, 0)$ as to $\zeta S_a(\theta, 0)$. The degree of unpredictability of the selection procedure is at a maximum for $\theta=\pi/2$ when $S_a$ contains equal numbers of $a\;$s and $\cancel a\;$s, and at a minimum when $\theta=0$ or $\pi$ (where $\zeta$ has no impact on the bit string). 

As discussed in the main body of the text, for computational convenience, the values at which $\cos^2 \theta/2$ is defined can be extended into the reals, and the $M$th roots of unity can be considered embedded into the complexes. Here it is claimed that such extensions result in the legendary conceptual problems of quantum theory. 

\subsection{Two Qubits}

In quantum theory, the general tensor-product form for a 2-qubit Hilbert state is given by
\be
\label{2qubit3}
|\psi_{ab}\rangle= \gamma_0 |a\rangle |b \rangle + \gamma_1 e^{i \chi_1} |a\rangle |\cancel{b} \rangle + \gamma_2 e^{i \chi_2}|\cancel{a}\rangle |b \rangle + \gamma_3 e^{i \chi_3}|\cancel{a}\rangle |\cancel{b} \rangle, 
\ee
where $\gamma_i, \chi_i \in \mathbb R$ and $\gamma_0^2+\gamma_1^2+\gamma_2^2+\gamma_3^2=1$. Equation (\ref{2qubit3}) can be written in two equivalent forms. The first is 
\be
\label{2qubit1}
|\psi_{ab}\rangle=
\cos \frac{\theta_1}{2}|a\rangle
|\psi_b(\theta_2, \phi_2)\rangle
+e^{i\phi_1} \sin \frac{\theta_1}{2} |\cancel{a}\rangle |\psi_b(\theta_3,\phi_3)\rangle
\ee
where
\begin{align}
\label{gamma}
\gamma_0&=\cos\frac{\theta_1}{2}\cos\frac{\theta_2}{2} \ \ \ \ 
\gamma_1=\cos\frac{\theta_1}{2} \sin\frac{\theta_2}{2} \ \ \ \ 
\gamma_2=\sin\frac{\theta_1}{2}\cos\frac{\theta_3}{2}  \ \ \ \ 
\gamma_3=\sin\frac{\theta_1}{2}\sin\frac{\theta_3}{2}  \nonumber \\
\chi_1&=\phi_2\ \ \ \ 
\chi_2=\phi_1 \ \ \ \ \
\chi_3=\phi_1+\phi_3 
\end{align}
The second is 
\be
\label{2qubit2}
|\psi_{ab}\rangle=
\cos \frac{\theta_6}{2} |\psi_a(\theta_4, \phi_4)\rangle  |b\rangle+ e^{i \phi_6} \sin \frac {\theta_6}{2} 
|\psi_a(\theta_5, \phi_5)\rangle |\cancel b\rangle
\ee
where
\begin{align}
\label{firsttosecond}
\cos \frac{\theta_1}{2} \cos \frac{\theta_2}{2} &= \cos \frac{\theta_4}{2} \cos \frac{\theta_6}{2} \nonumber \\
\sin \frac{\theta_1}{2} \cos \frac{\theta_3}{2} &= \sin \frac{\theta_4}{2} \cos \frac{\theta_6}{2} \nonumber \\
\cos \frac{\theta_1}{2} \sin \frac{\theta_2}{2} &= \cos \frac{\theta_5}{2} \sin \frac{\theta_6}{2} \nonumber \\
\sin \frac{\theta_1}{2} \sin \frac{\theta_1}{2} &= \sin \frac{\theta_5}{2} \cos \frac{\theta_6}{2} \nonumber \\
\phi_1&=\phi_4 \nonumber \\
\phi_2&=\phi_6 \nonumber \\
\phi_1+\phi_3&=\phi_5+\phi_6 \nonumber \\
\end{align}

\begin{figure}
\centering
\includegraphics[scale=0.4]{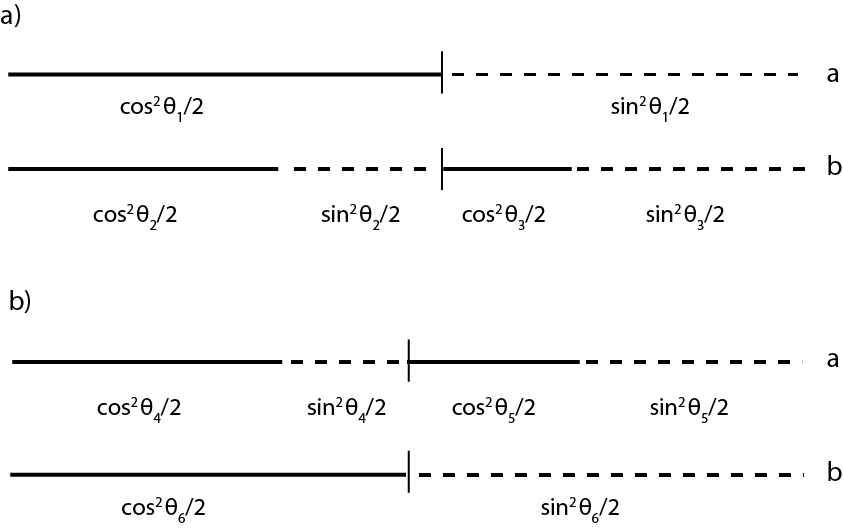}
\caption{A schematic illustration of the injective correspondence between (\ref{twobs}) and (\ref{2qubit3}) where phase angles are set to zero. In both a) and b) the top line refers to $S_a$, the bottom line to $S_b$. Solid lines illustrate sub-strings of bits which are either $a\;$s or $b\;$s and ordered differently in a) and b). Dashed lines refer to sub-strings of elements which are either $\cancel a\;$s or $\cancel b\;$s. Hence, for example, panel b) indicates that the first $2^N\; \cos^2 \theta_4/2\; \cos^2\theta$ elements of $S_a$ and $S_b$ are $a\;$s and $b\;s$ respectively. Despite different ordering, panels a) and b) are equivalent in terms of correlations between $a$ and $b$ (and other pairs of) elements because of the relations (\ref{firsttosecond}). Panel a) shows the correspondence between (\ref{twobs}) and (\ref{2qubit1}). Panel b) shows the correspondence between (\ref{twobs}) and (\ref{2qubit2}).  (\ref{2qubit1}) and (\ref{2qubit2}) are equivalent forms of the tensor product Hilbert state (\ref{2qubit3}). The role of phase angles is described in the text.}
\label{tensor}
\end{figure}

Let us start by assuming all $\phi_i=0$. Then in invariant set theory, a pure Hilbert state of the form (\ref{2qubit1}) is considered an uncertain selection of some particular pair of elements $\{a_i, b_i\}$ from the two bit strings
\begin{align}
\label{twobs}
S_a&=\{a_1 \; a_2 \ldots a_{2^N}\} \nonumber \\
S_b&=\{b_1 \; b_2 \ldots b_{2^N}\} 
\end{align} 
where $a_i \in \{a, \cancel a\}$, $b_i \in \{b, \cancel b\}$, where all $\cos^2 \theta_i/2$ are finitely describable, and where, as above, the probability of selecting $\{a_i, b_i\}$ is independent of $i$. Using the form of (\ref{2qubit1}) and Fig \ref{tensor} as guidance, these bit strings are defined as follows. With reference to the first line in Fig \ref{tensor}a, the first $2^N \cos^2 \theta_1/2$ elements of $S_a$ are $a\;$s, and the rest $\cancel a\;$s. With reference to the second line in Fig \ref{tensor}a, the first $2^N \cos^2 \theta_1/2 \; \cos^2 \theta_2/2$ elements of $S_b$ are $b\;$s; the next $2^N \cos^2 \theta_1/2 \; \sin^2 \theta_2/2$ elements are $\cancel b\;$s; the next $2^N\sin^2 \theta_1/2 \; \cos^2 \theta_3/2$ elements are $b\;$s and the final $2^N\sin^2 \theta_1/2 \; \sin^2 \theta_3/2$ elements are $\cancel b\;$s. Based on this we can assert that the uncertain selection $\{a_i, b_i\}$ corresponds to the Hilbert vector (\ref{2qubit1}) with $\phi_1=\phi_2=\phi_3=0$. 

Now if we use the relationships (\ref{firsttosecond}) it is easily seen that the probability of selecting $a$ from $S_a$ and $b$ from $S_b$ in Fig \ref{tensor}a is the same as in Fig \ref{tensor}b (as are the probabilities for any other pairs of elements). These probabilities are the same as implied by the Born rule on (\ref{2qubit3}) with $\chi_i=0$. For example, the probability that $a_i=a$ and $b_i=\cancel b$ is equal to $\cos^2 \theta_1/2\; \sin^2 \theta_2=\cos^2 \theta_5/2\; \sin^2 \theta_6$ which is equal to  $\gamma^2_1$ from (\ref{gamma}) and (\ref{firsttosecond}). 

The three phase degrees of freedom are introduced through the cyclical permutation operators $\zeta$. Note that we can cyclically permute the two strings $\{a_i\}$ and $\{b_i\}$ together without affecting the correlations between $S_a$ and $S_b$. Similarly, from Fig \ref{tensor}a, one can cyclically permute the first $\cos^2 \theta_1/2$ elements of $S_b$, or the final $\sin^2 \theta_1/2$ elements of $S_b$, without affecting the correlations between $S_a$ and $S_b$. Similarly, from Fig \ref{tensor}b, one can cyclically permute the first $\cos^2 \theta_6/2$ elements of $S_a$, or the final $\sin^2 \theta_6/2$ elements of $S_a$, without affecting the correlations between $S_a$ and $S_b$. As before, it  is important to note that because of the linkage to cyclical permutations, the corresponding phase angles must be rational multiples of $\pi$. 

Now consider the special case where $\theta_1=\pi/2$, $\theta_2=\pi-\theta_3$ (or equivalently, $\theta_6=\pi/2$, $\theta_5=\pi-\theta_4$). Then the  correlation between $S_a$ and $S_b$ is equal to $-\cos \theta_2=-\cos \theta_4$ and consistent with quantum theoretic correlations of measurement outcomes on the Bell state
\be
\frac{|a\rangle|b\rangle+ | \cancel{a} \rangle | \cancel{b} \rangle}{\sqrt 2}
\ee
where $\theta_2=\theta_4$ denotes the relative orientation of Alice and Bob's measurement apparatuses, with finitely describable cosine. The finite describability of the cosine of the relative orientation of Alice and Bob's measuring apparatus is fundamental in invariant set theory's account of the Bell Theorem.   

It is a simple matter to use the principle of induction to extend the construction described in this Section to link general $m$-qubit tensor-product Hilbert states (with rational squared amplitudes) to families of $m$ (generically partially correlated) bit strings. 

\section{CHSH}
\label{CHSH}

Let $x=0$, $x=1$, $y=0$, $y=1$ represent four distinct, randomly chosen points on $\mathbb S_2$ (corresponding to four distinct randomly chosen directions in physical space). Join together each pair of points by a great circle. Let $\theta_{xy}$ denote the angular distance between any $x$ point and any $y$ point.

Let $\alpha$ denote the relative angle between the two possible measuring directions $x=0$ and $x=1$ of Alice's measuring apparatus, and let $\beta$ denote the corresponding angle between $y=0$ and $y=1$ of Bob's apparatus. Suppose Alice chooses $x=0$. It is always possible for Alice to send a particle which she has just measured along direction $x=0$, back into the instrument to be measured in the $x=1$ direction. This corresponds to a simple (single-qubit) measurement where the input state has been prepared along $x=0$ and the measurement taken along $x=1$ (or \emph{vice versa} if Alice instead chose $x=1$). This means, according to invariant set theory, it must be the case that $\cos \alpha$ and $\cos \beta$ are finitely describable. 

Now consider a spherical triangle comprising any three of the four distinct points $x=0$, $x=1$, $y=0$ and $y=1$. For definiteness, consider the triangle whose vertices are $x=0$, $x=1$ and $y=0$. The cosine of the angular distance the side joining $x=0$ and $x=1$ is equal to $\cos \alpha$ and therefore is finitely describable. Suppose Alice and Bob choose $x=0$ and $y=0$ so that $\cos \theta_{00}$ is also finitely describable. Then, according to the cosine rule for spherical triangles
\be 
\label{cosinerule}
\cos \theta_{01}=\cos \theta_{00} \cos \alpha + \sin \theta_{00} \sin \alpha \cos \gamma
\ee
where $\gamma$ is the angle subtended at $x=1$. If $\gamma=0$ then we have the usual co-planar arrangement of angles. By the theory of Pythagorean triples, if $\cos \theta_{00}$ is finitely describable, then $\sin \theta_{00}$ is not finitely describable (there are no Pythagorean triples $\{a,b,c\}$ where $c$ is a power of $2$). Similarly, if $\cos \alpha$ is finitely describable, then $\sin \alpha$ is not. Since $\theta_{00}$ and $\alpha$ are essentially independent angles (reflecting the independence and finite precision of Alice and Bob's apparatuses), we can assume that the product $\sin \theta_{00} \sin \alpha$ is not in general finitely describable. Hence $\cos \theta_{01}$ is the sum of a term which is finitely describable, and a term which isn't. Hence $\cos\theta_{01}$ is not finitely describable. With $\gamma$ a further independently chosen angle, one can assume that this result holds for non-zero $\gamma$ too. 

Hence, with $z=x+y \mod 2$, in any CHSH experiment the sample space $\Lambda_{z=0}$ of supplementary variables from which $C(0,0)$ and $C(1,1)$ are computed (based on finitely describable $\cos \theta_{00}$ and $\cos \theta_{11}$) must be disjoint from the sample space $\Lambda_{z=1}$ from which $C(0,1)$ and $C(1,0)$ are computed (based on finitely describable $\cos \theta_{01}$ and $\cos \theta_{10}$). Hence, the position of the four points in Fig \ref{fig:CHSH}a is not identical to that in Fig \ref{fig:CHSH}b. This is not physically inconsistent because the measurement orientations can only be set to some finite precision. 

\section{The Dirac Equation in the Singular Limit $N=\infty$}
\label{Dirac}

In this Appendix we discuss the simplest form of the Dirac equation
\be
\label{Dirac1}
i\hbar \gamma_{\mu}\; \partial_{\mu}\psi - mc \psi = 0
\ee
for a particle of mass $m$ in order to describe dynamical evolution between preparation and measurement in invariant set theory. 

As is well known, the Dirac 4-spinor can be written as 2 Weyl 2-spinors. We will associate the latter with fields of $2^N$-element bit strings $S_a(\mathbf x, t)$, $S_b(\mathbf x, t)$. 
Consider a frame where the particle is at rest at $\mathbf x=0$. Let 
\begin{align}
\label{evolution1}
S_a(0, t) &= \zeta^n S_a(0,0) \nonumber \\
S_b(0, t) &= \zeta^{-n} S_b(0,0)
\end{align}
where
\be
\label{n}
n= \frac{2^N mc^2}{2 \pi \hbar} \  t
\ee
From (\ref{zetan}) we can write
\be
\zeta^n \equiv e^{2 \pi i n/2^{N}}= e^{i \omega t}
\ee
where $\hbar \omega=mc^2$. As before, the set $\{e^{i \omega t}\}$ of time evolution operators is isomorphic to the multiplicative group of complex phases $\phi$ where $\phi / 2\pi$ is describable by $N-1$ bits. With $E=mc^2$, then 
\be
\label{E}
E=\hbar \omega
\ee
and with $E$ a source of space-time curvature, this iconic equation of quantum mechanics can be interpreted is a manifestation of  (\ref{generalisedequation}), linking space-time geometry to the periodic state-space geometry of $I_U$. A key property of (\ref{evolution1}) is the granularity of time. From (\ref{n}), the unit $\Delta t$ of granularity is given by
\be
\Delta t = \frac{2 \pi \hbar}{2^N mc^2}
\ee
the consequences of which will be developed elsewhere. 

Writing 
\be
\psi(t)=\bp S_a(0, t) \\ S_b(0, t) \ep
\ee
then the evolution equation (\ref{evolution1}) can be written as
\be
\label{evolution2}
\psi(t)=
\bp e^{i \omega t} & 0 \\ 0 & e^{-i \omega t} \ep
\psi(0)
\ee
For any finite $N$, $\{e^{i \omega t}\}$ is not closed under addition (see (\ref{noadd})). As discussed in the main body of the text, this is considered a desirable property of invariant set theory, making it counterfactually incomplete. In the singular limit $N=\infty$, $e^{i \omega t}$ can be identified with the familiar complex exponential function, in which case, $e^{i \omega \Delta t} \approx 1+ i \omega \Delta t$ for small $\Delta t$ and
\be
i\partial_t \ e^{i \omega t}+ \omega e^{i \omega t}=0 \nonumber
\ee 
Because the limit is singular, the derivative is undefined for any finite $N$, no matter how big. In this sense, the Dirac equation for a particle at rest,
\be
\label{Dirac1}
i\hbar \gamma_0\; \partial_t \psi - mc^2 \psi = 0
\ee
corresponds to the singular limit of (\ref{evolution1}) at $N=\infty$. In (\ref{Dirac1}), $\psi$ is to be considered some abstract but computationally powerful `wavefunction', lying in complex Hilbert Space. From the perspective of invariant set theory, (\ref{evolution1}) and (\ref{n}) should be considered more fundamental, but less computationally powerful, than (\ref{Dirac1}).

In a non-rest frame, we must generalise (\ref{evolution1}) to include spatial variations and the full set of Dirac gamma matrices are needed to be relativistically invariant. Quantities such as 
\be
\gamma_i
\bp S_a\\ S_b\ep = 
\bp 0 & \sigma_i \\ -\sigma_i & 0 \ep
\bp S_a\\ S_b\ep
=\bp \sigma_i S_b \\ -\sigma_i S_a \ep
\ee
are straightforwardly defined by considering the Pauli matrices as operators on $S_a$ and $S_b$. For example
\begin{align}
\sigma_2 S_b &= 
\bp 0 & -i \\ i & 0 \ep
\{b_1\;b_2\ldots b_{2^{N-1}}\} \|
\{b_{2^{N-1}+1}\; b_{2^{N-1}+2}\ldots b_{2^N}\} \\
&=
\zeta^{2^{N-3}}\{\cancel b_{2^{N-1}+1}\; \cancel b_{2^{N-1}+2}\ldots \cancel b_{2^N}\} \|
\zeta^{2^{N-3}}\{b_1\;b_2\ldots b_{2^{N-1}}\}
\end{align}
where $\|$ denotes the concatenation operator. Full details will be given elsewhere. 

\section{The PBR Theorem}
\label{PBR}

The recent PBR theorem \cite{Pusey} is a no-go theorem casting doubt on $\psi$-epistemic theories (where the quantum state is presumed to represent information about some underlying physical state of the system). Unlike CHSH where Alice and Bob each choose measurement orientations A or B, here Alice and Bob, by each choosing $0$ or $1$, prepare a quantum system in one of four input states to some quantum circuit: $|\psi_0\rangle |\psi_0\rangle$, $|\psi_0\rangle |\psi_1\rangle$, $|\psi_1\rangle |\psi_0\rangle$ or $|\psi_1\rangle |\psi_1\rangle$, where
\begin{align}
|\psi_0\rangle&=\cos \frac{\theta}{2} |0\rangle + \sin \frac{\theta}{2} |1\rangle \nonumber \\
|\psi_1\rangle&=\cos \frac{\theta}{2} |0\rangle - \sin \frac{\theta}{2} |1\rangle \nonumber
\end{align}
In addition to the parameter $\theta$, the circuit contains two phase angles $\alpha$ and $\beta$; as discussed below, the phase angle $\alpha$ most closely plays the role of the phase angle $\phi$ in the Mach-Zehnder interferometer in Section \ref{interfere}. The output states of the quantum circuit are characterised as `$\mathrm{Not}\; 00$', `$\mathrm{Not}\; 01$', `$\mathrm{Not}\; 10$' and `$\mathrm{Not}\; 00$'.  The $\alpha$ and $\beta$ are chosen to ensure that (according to quantum theory), if Alice and Bob's input choices are $\{IJ\}$ where $I,J\in\{0,1\}$, then the probability of `$\mathrm{Not}\; IJ$' is equal to zero. However, if physics is governed by some underpinning $\psi$-epistemic theory, then, so the argument goes, at least occasionally the measuring device will be uncertain as to whether, for example, the input state was prepared using $00$ and $01$ and on these occasions it is possible that an outcome `$ \mathrm{Not} \; 01$' is observed when the state was prepared as $01$, contrary to quantum theory (and experiment). How does Invariant Set Theory, which is indeed a $\psi$-epistemic theory, avoid this problem?

Working through the algebra, it is found that the probabilities of various outcomes are trigonometric functions of $\alpha-\beta$, $\alpha-2\beta$, $\beta$ and $\theta$. For example, if Alice and Bob chose $00$, then, according to quantum theory (and therefore experiment), the probability of obtaining the outcome `$\mathrm{Not}\; 01$' is equal to 
\be
\label{XXX}
X= \cos^4 \frac{\theta}{2} + \sin^4 \frac{\theta}{2} +2 \cos^2 \frac{\theta}{2} \sin^2 \frac{\theta}{2}\cos (\alpha-2\beta) \ne 0
\ee
On the other hand, if Alice and Bob chose $01$, then the probability of obtaining the outcome `$\mathrm{Not}\; 01$' would be equal to 
\be
\label{ZZZ}
Z=X-4 \cos^2 \frac{\theta}{2} \sin^2 \frac{\theta}{2}-4 \cos^3 \frac{\theta}{2} \sin \frac{\theta}{2}\cos(\alpha-\beta)-4 \cos \frac{\theta}{2} \sin^3 \frac{\theta}{2}\cos\beta=0
\ee
The key point is that $X$ contains the trigonometric term $\cos(\alpha-2\beta)$, whilst $Z$ contains the terms $\cos(\alpha-\beta)$ and $\cos \beta$. Now one can clearly find values for $\alpha$, $\beta$ and $\theta$ such that $X$ is described by $N$ bits. That is to say, for large enough $N$, Invariant Set Theory can predict the quantum theoretic probability of outcome `$\mathrm{Not}\;01$' when Alice and Bob chose 00. However, in general it is impossible to find values for these angles such that $X$ and $Z$ are \emph{simultaneously} describable by $N$ bits. The number-theoretic argument is exactly that used to negate the Bell Theorem. For example, if $\cos (\alpha-2\beta)$ and $\cos \beta$ are describable by a finite number of bits, then $\cos(\alpha-\beta)=\cos(\alpha-2\beta) \cos \beta+ \sin(\alpha-2\beta) \sin \beta$ is not. This means that if in reality Alice and Bob chose $00$ in preparing a particular quantum system and the outcome was `$\mathrm{Not}\; 01$' , then there is no counterfactual world on $I_U$ where Alice and Bob chose $01$ in preparing the same quantum system, and the outcome was again `$\mathrm{Not}\; 01$'. That is to say, it is not the case that $Z=0$ for this counterfactual experiment - rather, $Z$ is undefined. Conversely, if in reality Alice and Bob chose 01, then there exist values for $\alpha$, $\beta$ and $\theta$ such that $Z$ is described by $N$ bits and equal to zero (to within experimental accuracy) and `$\mathrm{Not}\; 01$' is not observed. 

Let $\{\alpha_X, \beta_X, \theta_X\}$ denote a set of angles such that $X$ is describable by $N$ bits, and $\{\alpha_Z, \beta_Z, \theta_Z \}$ a set of angles such that $Z$ is describable by $N$ bits, i.e. these correspond to experiments on $I_U$. Now, as before, we can find values such that the differences $\alpha_Z-\alpha_X$, $\beta_Z-\beta_X$ and $\theta_Z-\theta_X$ are each smaller than the precision by which these angles can be set experimentally. Hence, Invariant Set Theory can readily account for pairs of experiments performed sequentially with seemingly identical parameters, the first where Alice and Bob choose $00$ and the outcome is sometimes `$\mathrm{Not}\; 01$', and the second where Alice and Bob choose $01$ and the outcome is never `$\mathrm{Not}\; 01$'. That is to say, the Invariant Set Theoretic interpretation of the PBR quantum circuit reveals no inconsistency with experiment. Even though Invariant Set Theory is $\psi$-epistemic, the holistic structure of the invariant set $I_U$ ensures that the measuring device will never be uncertain as to whether, for example, the input state was prepared using $00$ and $01$.

It was shown above that Invariant Set Theory evades the Bell theorem by violating the Measurement Independence assumption. Here it has been shown that Invariant Set Theory evades the PBR theorem by violating an equivalent Preparation Independence assumption. As before, this does not conflict at all with the experimenter's sense of free will. Neither does it imply fine-tuning with respect to the physically relevant $p$-adic metric on $I_U$. 

\end{document}